\documentclass{aa}
\usepackage{graphics}

\begin{document}

\thesaurus{06(08.19.4; 02.05.2; 02.14.1)}
\title{Nuclear properties in early stages of stellar collapse.}



\author{F. K. Sutaria \inst{1}
 \and A. Ray \inst{2,3}
 \and J. A. Sheikh \inst{2,4}
 \and P. Ring \inst{4}
}
\offprints{A. Ray, akr@tifr.res.in}

\institute{Inter University Center for Astronomy and Astrophysics,
Pune 411 007, India
\and Tata Institute of Fundamental Research, Mumbai 400 005, India
\and Lab for High Energy Astrophysics,
NASA/Goddard Space Flight Center, Greenbelt, MD 20771, U.S.A.
\and Physik-Department, Technische Universitat Munchen,
85747, Garching, Germany}
\date{Received 31 July, 1998; Accepted 13 April, 1999 }
\mail{Tata Institute of Fundamental Research,
Mumbai 400 005, India}
\titlerunning{Nuclear properties in stellar collapse: RMF}
\authorrunning{F. K. Sutaria et al.}

\maketitle

\vspace{1cm}
\begin{abstract}
\label{sec:abstract}
The spectroscopy of electron capture neutrinos emitted from nearby 
pre-supernova collapsing stars before the neutrino trapping sets in, 
can yield useful information on the physical conditions and on the
nuclear composition of the core. The neutrino spectrum depends on the  
thermodynamic conditions of the core, the nuclear abundances, the 
lepton fractions and relevant nuclear properties.
In the pre-trapping core, the density ranges from $0.1 - 100 
~10^{10}$ g/cm$^3$ and the temperature 
from $0.2 - 1.5$ MeV. The nuclear abundances can be obtained 
under the assumption of Nuclear Statistical Equilibrium (NSE).
The nuclear abundances as well as the electron capture rates are 
thus determined, among other things, by the nuclear binding energies 
and the free nucleon chemical potentials. Because shell 
and pairing effects persist strongly up to temperatures of $\simeq 0.5$ MeV, 
any equation of state (EOS) relevant to this phase of the collapse must
reproduce well the zero temperature nuclear properties and it must
show a smooth transition to the known high temperature and high
density limits. In this work we use the microscopic Relativistic 
Mean Field (RMF) theory based on a Lagrangian with non-linear 
self-interactions of the $\sigma$-field for the neutron-rich nuclei
of interest in the $f-p$ shell to determine nuclear chemical potentials.
We compare these results with those computed from an EOS calculated
with the macroscopic liquid drop model. We also discuss extensions
to finite temperature and we incorporate nuclear lattice effects 
into the microscopic calculations. 


\end{abstract}   

\section {Introduction}
\label{sec: Intro}

The collapse of the core of a massive star is believed to lead to 
the explosion of a type II supernova. So far, however, numerical 
simulations of such a collapse have not been able to demonstrate
convincingly the explosions observed in nature. Such simulations 
require many detailed and complex physical inputs, such as the 
equation of state of dense and warm stellar matter, the hydrodynamics 
of the shock wave, or the energy transport by neutrinos. The calculation 
of the equation of state of the stellar core material at densities below the
nuclear saturation densities ($ 10^9 < \rho < 10^{13} $ g/cm$^3$) 
constitutes, technically, one of the most computer intensive parts, because 
it requires a high accuracy due to its influence on the early part of the 
supernova development (see e.g. Cooperstein \& Baron 1990).

Initially, at densities of $10^9$ g/cm$^3$ the core is composed 
of heavy nuclei immersed in a electrically neutral plasma of electrons, 
with a small fraction of drip neutrons and an even smaller fraction of 
drip protons. Nevertheless, the overall density being much lower than 
nuclear matter density, the average volume available to a single nucleus 
is much greater than that of the nuclear volume. Under such circumstances
the equation of state can be approximated by that of a Boltzmann gas of 
heavy nuclei, and drip particles in Nuclear Statistical Equilibrium (NSE), 
together with a highly degenerate, electrically neutralizing gas of 
non-interacting electrons. This state of affairs persists up to densities
of $~\sim 10^{12} ~\rm g/cm^3$, where  the ratio of the nuclear radius 
($R_N \sim A^{1/3}$) to the Coulomb interaction radius ($R_c \sim (A/ 
\rho)^{1/3}$) given by $R_N / R_c \simeq 10^5 \rho^{1/3}$ is no longer 
small $\ll 1$ and the nuclei start interacting with each other by means 
of Coulomb lattice effects. Finally, at values around 
$ 10^{13} ~\rm g/cm^3$ the density of the "vapor" of free nucleons 
surrounding the droplets of nuclear matter is too high to justify the 
treatment of nuclei as a Boltzmann gas of {\em non-interacting} particles  
(see e.g. Hillebrandt (1994) and references therein).

The equation of state of moderately neutron-rich matter at intermediate 
temperature and density can substantially influence the energy spectrum 
and the flux of the early phase neutrinos from electron capture on nuclei 
and free protons. Should the neutrinos from the collapse
phase be detected from a nearby supernova 
in an underground neutrino detection experiment such as the Sudbury 
Neutrino Observatory (SNO) or the Super-Kamioka (SK), such spectra 
could be possibly measured. This will be possible only under
the assumptions, that such a supernova explosion will occur at a distance
of approximately 1 kpc and that these new generation of neutrino detectors with
increased sensitivity are available at that time (see Sutaria \& Ray (1997),
hereafter SR). 
The neutrinos from the early phase of the collapse, before they are 
trapped by dense overlying matter are especially interesting for several 
reasons. They bear information directly related to the conditions
under which they are produced (SR). 
The pre-trapping phase of the emission of core-neutrinos continues until 
the overlying density grows to about $10^{12}$ g/cm$^{3}$. The emitted 
neutrino spectrum depends upon the individual $Q$-values  $Q_i$ of the 
nuclei involved in the electron-capture reaction. Both these $Q$-values 
as well as the overall abundance of particular nuclei depend upon the 
binding energies of these nuclei. As the electron capture on protons 
has different neutrino emission characteristics than the capture on nuclei, 
the emitted spectra will be different for different admixtures of these 
components. The nuclear equation of state (EOS) controls the heights and the
centroids of the neutrino spectrum such as the example shown in Fig \ref{fig1:
25msmhv2}. This example was computed using the EOS, in
particular the expressions for nucleonic chemical potentials as given by 
Fuller (1982).

During the collapse, it depends also on the entropy of the core 
whether most of the electron captures occur on free protons or on 
heavy nuclei. At relatively low entropies (e.g. $S/k_B \approx 1$)
captures on heavy nuclei dominate the total rate. Such entropies 
of the core do occur for stars of main sequence mass 10 - 25 $M_{\sun}$. 
As the entropy connected with nuclear excitations and the entropy of the 
free nucleons form a significant part of the total entropy of the stellar 
core, this forms another reason for evaluating the nuclear properties 
of the stellar core with accuracy. The relevant temperature range of the 
stellar core is approximately 0.2 to 1.5 MeV at such low entropies for 
a density range of $10^9$ -- $10^{12}$ g/cm$^3$.

Investigations on the equation of state of supernova matter have 
by now spanned several decades (Bethe et al. 1979, hereafter BBAL;
Lamb et al. 1978, hereafter LLPR78; Lattimer et al. 1985, hereafter LPRL85;
Lamb et al. 1983, hereafter LLPR83).
At the same time, ever since the discovery of neutron stars as pulsars 
more than thirty years ago, work on the equation of state of ``cold, 
catalyzed" matter, i.e. matter at zero temperature and under thermodynamic 
equilibrium, has been pursued by a number of scientists (see 
Pethick \& Ravenhall 1995 and references therein). 
The equation of state of hot 
supernova matter has been in many ways an extension of the early work 
done on the EOS of neutron star matter at zero temperature 
(Baym et al. 1971a (BPS); Baym et al. 1971b (BBP)).
Since these investigations were based on the 
liquid drop mass formulae of Myers \& Swiatecki (1966), 
the calculations of the equations of state for supernova have 
also followed largely this semi-empirical approach. More recent 
attempts of the EOS of supernova matter by Swesty et al. (1994) 
are based on the finite temperature, compressible 
liquid droplet model which attempts to take into account the 
nucleon-nucleon interaction in analogy to the density 
dependent Skyrme type energy functionals. On the other hand, there are
microscopic investigations based on the Hartree-Fock or on the 
Hartree-Fock-Bogoliubov scheme with effective nucleon-nucleon 
interactions, such as that of Negele \& Vautherin (1973) above
the neutron drip point or that of Haensel et al. (1989)
below the neutron drip point. 

However, as far as pre-supernova conditions are concerned, the liquid-drop
(or droplet) based versions of the EOS cannot be applied so easily because 
in this temperature range (0.2 to 1.5 MeV) shell and pairing effects
dominate the nuclear structure. It is only at temperatures beyond 
$\sim 1.0 {~\rm to}~ 1.5$ MeV, that nuclei can be dealt within the 
liquid drop approximation. Thus an equation of state which 
is used to calculate nuclear properties in the pre-supernova stage 
should take into account ``microscopic" shell  and pairing effects. 
At higher temperatures and densities, well after neutrino trapping at 
$10^{12} ~\rm g/cm^3$ they should smoothly go over into the 
high-temperature high-density EOS based on the liquid drop model 
of nuclear matter. Furthermore the computation of the EOS by
theoretical methods  needs to be supplemented by experimental
data and their trends.

In this context, we note the work of El Eid \& Hillebrandt (1980) 
(hereafter EH)
which computed the equation of state of supernova matter and made use of
the so called droplet model (DM) and its extension.
The DM model was an early attempt which 
went beyond the conventional liquid drop model (LDM) also due
to Myers \& Swiatecki
(1966) with the inclusion of a shell correction term due to 
von Groote et al. (1976). The LDM was used for the EOS computations 
of much of the subsequent works mentioned below.
The Droplet Model is an analytical extrapolation of the Thomas Fermi model.
Von Groote et al. also gave a new fit to the DM parameters to all
nuclear masses known at that time.
In this paper, we present a microscopic calculation based on a suitably 
generalized Relativistic Mean Field theory adapted to the stellar 
collapse problem (we refer also to Sutaria (1997), Ray et al. 
(1999) and Sutaria et al. (1997)). This automatically accounts
for the nucleon-nucleon interactions and for shell and pairing
effects, and as we show later reproduces the currently known nuclear masses
with substantial accuracy. A microscopic approach like this is
expected to be more reliable for nuclei relevant in the astrophysical context,
which are often very neutron rich and are thus far from the valley of
$\beta$-stability.

In section \ref{sec: precEOS} we summarise the existing equations 
of state based on the liquid drop approach; this helps to 
establish the notation and to compare with previous approaches. 
In section \ref{sec: RMFapp}  we describe the Relativistic Mean Field theory 
calculations as adapted to the supernova context. In this section 
we also present the results of these investigations. In section \ref{sec: Comp} 
we compare our results with those of liquid drop based models and
compare with the earlier work of Cooperstein and Baron (1990)
and references therein. 
In section \ref{sec: FiniteTemp} we describe the generalization to 
finite temperature and discuss the modification 
of our results due to low but finite temperature effects.
In the concluding section \ref{sec: Summ} we  make a comparison with the
results of EH and ours at the initial stage of stellar collapse
and summarize our conclusions.

\section{Suitable Equations of State for each Stage of the Collapse}
\label{sec: precEOS}

The initial efforts to compute the EOS, and indeed most of the literature 
available so far on the EOS of supernovae concentrate mainly on matter 
at densities beyond $10^{13}$ g/cm$^3$. Most of them are based on the 
liquid drop model of the nucleus, as the initial work of BBAL, 
LLPR and their co-workers. Later 
finite temperature extensions of the existing EOS as for instance
the work of Baron et al. (1985) (hereafter BCK)
and Cooperstein (1985) 
took into account the nuclear compressibility.
The most sophisticated attempts in developing an equation of state at 
higher temperatures and higher densities, close to nuclear saturation 
densities, have been finite temperature Hartree Fock calculations, 
as those of Bonche \& Vautherin (1981). These Hartree-Fock 
approaches rely on the spherical approximation and on phenomenological 
models for the nuclear potentials, usually of Skyrme type, which are 
fitted carefully to properties of finite nuclei.  

As discussed in Sect. \ref{sec: Intro}, the composition of stellar 
matter at densities between $10^7$ and $10^{12}$ g/cm$^3$ and at
temperatures between $0.2$ and $1$ MeV can be approximated by a 
'soup' of nuclei including drip neutrons, protons, alpha-particles 
and an ensemble of heavy nuclei. The nuclear abundances in this 
pre-supernova stage  with a density $\rho \approx 10^7 $ g/cm$^3$ 
peak in two regions of the isotope table, one around the silicon group 
and the other around the iron group. The nuclear species in each of 
these groups remain in quasi-equilibrium among themselves but the 
groups of nuclei are not in quasi-equilibrium with each other at 
the lower densities (Hix \& Thielemann 1996). The fraction of 
drip nucleons remains low during the collapse until nuclei merge into 
a uniform nucleon sea, possibly through a bubble phase. This is because
the entropy per nucleon of the system remains close to 1 
(in units of k$_B$) implying a high degree of order in the system 
(BBAL).

In the macroscopic incompressible liquid drop approach to the cold EOS 
above the neutron drip (BBAL 1979; BBP 1971), 
the ensemble of heavy nuclei 
and drip nucleons is considered on a lattice of number density $n_N$ 
consisting of a single heavy nucleus with mass and charge numbers 
(A,Z) having a nuclear volume $V_N$ immersed in a sea of drip neutrons 
of density $n_n$ and electrons of density $n_e$. For such a system, the 
effective free energy density $E_{tot}$ at zero temperatures (the internal 
energy density of the system), is given by BBP:
\begin{eqnarray}
\lefteqn{ E_{tot}( A, Z, n_N, V_N, n_n) =} \nonumber \\
& & n_N(W_N + W_L) 
 +(1- V_N n_N)E_n(n_n) + E_e(n_e). 
\label{eq: Etot}
\end{eqnarray}
$W_L$ is the Coulomb lattice energy of the system, $(1-V_Nn_N)$ is the 
fraction of the total volume occupied by the neutron gas, 
$E_n$ is the energy density of the non-relativistic free nucleon gas
consisting primarily of drip neutrons
and $E_e(n_e)$ is the energy density of
the relativistic electrons.  

The lattice energy of the system, assuming a bcc lattice
structure, is expressed as $ W_L= -1.82 Z^2 e^2 /a $, 
where $a= (2/n_N)^{1/3}$ is the lattice constant.
The quantity $W_N$ is the nuclear matter energy of a nucleus in equilibrium
with the free neutron gas, and is given by BBP:
\begin{eqnarray}
 W_N & = & [(1-x)m_n + x m_p]c^2A + [W + W_{thick} + W_{exch}]A \nonumber \\
 & + & W_{surf}A^\frac{2}{3} + W_{c,0} Z^2 A^{-\frac{1}{3}} 
\label{eq: WN(BBP)}
\end{eqnarray}
where $x$ is the proton fraction of the heavy nucleus, $m_n$ and $m_p$
are neutron and proton masses, $W$ is the energy of bulk nuclear matter
per nucleon, $W_{surf}$ is the coefficient of the surface energy term
and $W_{c,0}$ is the nuclear Coulomb energy, which is taken to be the
Coulomb energy of a uniformly charged sphere with charge $Z$. There are two
corrections to the total Coulomb energy; a term due to the finite surface
thickness which is denoted by $W_{thick}A$, and another due to the proton 
exchange term $W_{exch}A$, but both these are generally negligible 
(BBP). The total nuclear energy $W_N$, including the corrections due 
to lattice energy term (because the system consists of a lattice of positively 
charged nuclei sitting in a sea of electrons) and the 2nd order corrections 
to this lattice energy (due to the non-zero Coulomb radius) is given in 
terms of  of the proton fraction $x$, the density in the nuclear interior 
$\rho_N$, the nuclear volume $V_N$ and the packing fraction i.e. 
the fraction of total volume occupied by nuclei ($u=\rho/\rho_N$), as 
(Bethe et al. 1983, hereafter BBCW):
\begin{eqnarray}
W_N( x, \rho_N, V_N, u) & = & W_{bulk} + 290x^2(1-x)^2 A^\frac{2}{3} \nonumber \\ 
 & & + \beta x^2 \rho_N^2
V_N^\frac{5}{3} (1 -\frac{3}{2} u^\frac{1}{3} + \frac{1}{2} u),
\label{eq: WN}
\end{eqnarray}
where the factor $\beta \simeq 3/5$, and $W_{bulk}$ is the energy of infinite 
nuclear matter without surface or Coulomb, but with asymmetry effects,
given in BBCW:
\begin{equation}
 W_{bulk} = [-16 + 29.3(1-2x)^2],
\label{eq: Wbulk}
\end{equation}
The above expression for $W_{bulk}$
is based on the simple {\em incompressible} liquid drop 
mass formula. The first term is the volume energy of the nucleus per nucleon. 
The second term represents the nuclear volume asymmetry energy and 
the coefficient of the volume asymmetry energy $S_v$. Here we use the
value $S_v= 29.3$ obtained in the literature from fits to the experimentally 
known nuclear masses (Swesty et al. 1994).

The mass number $A$ of the most dominant nuclear species is determined by 
a minimum of the free energy with respect to $A$ at constant $x$, $n_N A$, 
$n_N V_N$, $n_n$. This  leads to the conditions that the most stable nuclear 
system (for a given e$^-$-fraction $Y_e=x$, drip neutron density $n_n$, 
packing fraction $u$ and heavy nuclei density $\rho_N\simeq n_N A$) is 
that for which the surface energy is twice the Coulomb energy:
$W_{surf} = 2W_{coul}A$.

The chemical potential of the neutrons $\mu_n$ is defined as the change 
in the nuclear energy when a single neutron is added to the nucleus, 
keeping the number of protons constant. Therefore $\mu_n$ can be obtained 
from Eq. (\ref{eq: WN}) by
\begin{equation}
\mu_n = \frac{1}{V_N}\left(\frac{\partial W_N}{\partial \rho_N}
\right)_{x,V_N,u}
- \frac{x}{A}\left(\frac{\partial W_N}{\partial x}\right)_{\rho_N,V_N,u},
\label{eq: mun}
\end{equation}
and similarly, the difference of neutron and proton chemical potentials is 
obtained by:
\begin{equation}
\hat\mu = \mu_n - \mu_p = -\frac{1}{A}
\left(\frac{\partial W_N}{\partial x}\right)_{\rho_N,V_N,u}
\label{eq: muhat}
\end{equation}
Using Eqs. (\ref{eq: mun}) and (\ref{eq: muhat}) along with 
Eq. (\ref{eq: WN}) 
and the condition that the surface energy is twice the Coulomb energy  
for the "mean nucleus", one can obtain relations for $\mu_n$ and $\hat 
\mu$. Fuller (1982), quoting a private communication from Lattimer (1980)
gives  an expression for $\mu_n$ as :- 
\begin{eqnarray}
\mu_n & = & -16 + 125(0.5-x) -125(0.5-x)^2 \nonumber \\
      &   & -290\frac{x^2 (1-x)^2(3-7x)}{A^{1/3}2(1-x)}
\label{eq: Fullmun}
\end{eqnarray}
and for $\hat \mu$ as (BBAL) :-
\begin{eqnarray}
\hat \mu &=& 250(0.5 -x) - \frac{290}{A^{1/3}}  x^2 (1-2x)^2 
\left(\frac{1}{x} + \frac{2}{x}\frac{1-2x}{1-x}\right).
\label{eq: Fullmuh}
\end{eqnarray}

Note however the comments pertaining to these, 
discussed in the next section.
This model described in Eqs. (\ref{eq: Etot} - \ref{eq: Fullmuh})
(cited as BBAL/BBP-EOS later on) 
is based on an incompressible liquid drop.
It uses the surface energy term of Ravenhall et al. (1972) and
it assumes that the actual nuclear density can be approximated by 
the nuclear density at saturation. The relations for $\mu_n$ are taken
from Fuller (1982) and are meant for near-symmetric nuclear matter 
for which $x \simeq 0.5$. The correction terms in $\mu_n$ are
discussed in the next section. In Fig.\ref{fig2: ch5mn} we display 
the LDM expression of Eq. \ref{eq: Fullmun} for the chemical potential
of the neutrons in the chain of Mn-isotopes along with the values  
obtained form relativistic mean field calculations $\mu_n |_{RMF}$ which 
will be discussed later. Similar diagrams for the elements 
Fe, Co, Ni, Cu, Zn, Ga, and Ge can be found in Sutaria (1997). 
For the curve labeled BCK in Fig.\ref{fig2: ch5mn}, modifications were 
made to take into account the difference between the actual nuclear 
density and the saturation density in the compressible EOS developed 
by Baron et al. (1985) (BCK) and Cooperstein (1985) 
(cited as BCK-EOS lateron). 

The BCK-EOS differs from the BBAL/BBP-EOS in three points: (a) a term
is incorporated in the expression for $W_N/A$ which takes into account
the nuclear compressibility, (b) the actual nuclear density
$\rho_0$ may differ from the saturation nuclear density of infinite
nuclear matter $\rho_s$, (c)  $\rho_s$ is allowed to depend on the 
proton fraction $x$, where the functional form of this dependence 
($\phi(x)=\rho_s(x) /0.16 ~\rm fm^{-3} = 1-3(0.5 -x)^2$) was obtained 
by  fits to the density dependent HF calculations with the force SKM 
of Bonche \& Vautherin (1981). In the compressible model, the expression for 
$W_N$ has the form:
\begin{eqnarray}
W_N &=& -16 ~+~ 29.3\,(1-2x)^2 ~+~ \frac{1}{18} K (1- \theta)^2 
\nonumber\\
&&+~ 75.4\, x^2(1-x)^\frac{4}{3} \phi^{-\frac{1}{3}}\theta^{-\frac{1}{3}} 
\left(1-\frac{3}{2} u^\frac{1}{3}+\frac{1}{2} u\right)^\frac{1}{3}
\label{eq: Coop}
\end{eqnarray}
Where $K$ is the coefficient of nuclear incompressibility, 
$\theta = \rho_0/\rho_s$ is the ratio of the actual nuclear 
density $\rho_0$ to the saturation nuclear density $\rho_s$, 
and  $u$ is the packing fraction of the nuclei, such that 
$u=\rho/\rho_0$. The term dependent on the incompressibility 
was added in order to obtain an accurate description of the phase 
transition into the bubble region without producing any discontinuities 
in the pressure or the temperature variables in the numerical calculations 
simulating the core collapse. The low entropy liquid drop-based equations 
of state (hereby the LDM-EOS) all apply to warm dense matter. The approaches 
vary: BBAL use the phenomenological approach, 
which included many basic features of the microscopic calculations of LPRL.
BCK use the "top down" route, 
where the relevant nuclear parameters like bulk incompressibility, 
saturation density including its dependence on $Y_e$, on the nucleon 
effective mass, or the bulk asymmetry energy coefficients, etc. were taken 
as input quantities for the calculation of thermodynamic variables, 
rather than as output for a specific set of nuclear force parameters, 
as in the microscopic calculation. The BCK series of EOS were constructed 
for numerical applications in supernova hydrodynamics and therefore with 
physical and numerical simplicity as important considerations. 
The composition of warm dense matter is determined under the conditions 
of mass and charge conservation, by the pressure balance at the surface of a 
heavy nucleus surrounded by light nuclei like $\alpha$-particles and by 
a "vapour" of free protons and neutrons, and under the assumption that 
the strong and electro-magnetic reactions have reached nuclear statistical 
equilibrium (NSE) (Cooperstein \& Baron 1990). 
 
An explicit dependence on temperature was introduced in the BCK-EOS 
free energy ($F$) term by using the expression:
\begin{equation}
F= W_N - \frac{a}{A} \frac{m^*}{m} T^2
\label{eq: temp}
\end{equation} 
where $m^*/ m$ is the effective mass of the nucleons in the nucleus
and $a$ is level density parameter in the Fermi-gas model of the nuclei.

\section{The Relativistic Mean Field (RMF) Approach}
\label{sec: RMFapp}
\medskip
The Hartree-Fock (HF) or Hartree-Fock-Bogoliubov (HFB) methods of 
calculating nuclear properties depend on the knowledge of a suitable 
nuclear potential. Usually Skyrme or Skyrme-type forces are used for 
this purpose. In recent years the microscopic description of ground state
properties of finite nuclei has been attempted by a relativistic
field theory for the nuclear many-body problem. Reviews of this 
work have been given by Serot \& Walecka (1986) and Gambhir et al.
(1990).

Relativistic Mean Field Theory starts from an effective Lagrangian 
containing the nucleonic and mesonic degrees of freedom. It is a 
relativistic analogue of the concepts of density dependent Hartree-Fock 
calculations with Skyrme forces. Because of its proper treatment
of the spin-orbit splitting this model is expected to be more reliable 
than the non-relativistic models in predicting yet unknown properties 
of nuclei far from stability which are important in astrophysical situations. 
Also in some respects this method is simpler than the Skyrme type 
calculations, since the RMF method involves only local quantities 
such as local densities and fields. The binding energies and nuclear 
charge radii calculated by these methods agree well with the experimental
values (see e.g. Table \ref{tab1: ch-radii}). 
The density distributions of doubly magic spherical nuclei also 
agree well with electron scattering data. In addition, apart from the 
ground state properties of spherical nuclei, RMF reproduces the right 
ordering of the single particle spectra in adjacent odd mass nuclei, 
without any additional parameters. It also describes in a quantitative
way nuclear deformations and superdeformations. 

The microscopic RMF approach is used here  to calculate nuclear properties 
at zero temperature, both for isolated nuclei, an approximation which holds 
well at the end of the Si-burning stage as well as for the case with 
higher density but still zero temperature where the nuclei can be assumed 
to be distributed in a lattice, and where modifications due to Lattice 
Coulomb energy have to be taken into account. 

Within relativistic mean field theory (Sheikh et al. 1993) the nucleus is
described a an ensemble of nucleons described by Dirac spinors and
moving independently in self-consistently determined meson- and electro-magnetic
fields. The Lagrangian, describing this system is given by
\begin{equation}
{\cal L}~=~ {\cal L}_B ~+~ {\cal L}_M ~+~ {\cal L}_{BM}, 
\end{equation}
with
\begin{eqnarray}
{\cal L}_B &=& \overline\psi_i( i \gamma^{\mu}\partial_{\mu}-m_N) \psi_i,
\\
{\cal L}_M &=& \frac{1}{2} \partial^{\mu} \sigma \partial_{\mu} \sigma
-U(\sigma)
-\frac{1}{4}\Omega^{\mu \nu}\Omega_{\mu \nu} + 
\frac{1}{2}m_{\omega}^2\omega^{\mu}\omega_{\mu} \nonumber \\
 & & -\frac{1}{2}\tilde{R}^{\mu \nu}\tilde{R}_{\mu \nu} + 
\frac{1}{2}m_{\rho}^2\tilde{\rho}^{\mu}\tilde{\rho}_{\mu} 
-\frac{1}{2}F^{\mu \nu}F_{\mu \nu},
\end{eqnarray}
\begin{eqnarray}
{\cal L}_{BM} & = &  
-g_{\sigma}{\overline\psi_i}\psi_i\sigma 
-g_{\omega}{\overline \psi_i}\gamma^{\mu}\psi_i \omega_{\mu} \nonumber \\ 
& & -g_{\rho}{\overline\psi_i}\gamma^{\mu}\tilde{\tau}\psi_i \tilde{\rho}_{\mu} 
-e {\overline \psi_i} \gamma^{\mu} \frac{1- \tau_3}{2} \psi_i A_{\mu}.
\end{eqnarray}
$m_N$ is the nucleon mass, $m_{\sigma}$, $m_{\omega}$,  and $m_{\rho}$
are the masses of the $\sigma$-, $\omega$- and $\rho$-mesons,  
$g_\sigma$ $g_\omega$ and $g_\rho$ are the corresponding meson-nucleon
coupling constants. Tilde indicate the isospin degree of freedom
and $\Omega^{\mu\nu}$, $\tilde{R}_{\mu\nu}$, $F_{\mu\nu}$ are the
field tensors. $U(\sigma)$ is the non-linear potential for the 
$\sigma$ mesons. It takes into account the density dependence in 
a phenomenological way and it has the form (Boguta \& Bodmer 1977): 
\begin{equation}
U(\sigma)=\frac{1}{2} m_{\sigma}^2 {\sigma}^2 + \frac{1}{3}g_2 \sigma^3
+\frac{1}{4} g_3 \sigma^4
\end{equation}
For the three mass parameters $m_N$, $m_{\omega}$ and $m_{\rho}$ entering
the Lagrangian we use the experimentally known values. In general, the 
other parameters are obtained by fitting ground-state binding energies 
and charge radii of a few spherical nuclei. Several sets of such parameters
are given in the literature. The parameter set NL1 (Reinhard et al. 1986)
has turned 
out to be very successful in the valley of $\beta$-stability. Here we use 
the parameter set NL-SH, which was obtained by fitting the binding energy, 
the charge-radii and also the neutron-skin radii of a few spherical nuclei 
(Sharma et al. 1993).
In table \ref{tab1: ch-radii} we show that both the 
NL1 and the NL-SH parameter sets reproduce well the experimentally
known nuclear charge-radii (Nadjakov et al. 1994)
of Ni isotopes. We also compare (see Table \ref{tab7: NL3})
with results obtained from the more recently determined parameter set 
NL3 (Lalazissis et al. 1997), which is often used in the literature 
for the study of exotic nuclei.

While the set NL-SH reproduces well nuclear binding energies, we find that 
the incompressibility coefficient $K_0$ of bulk nuclear matter as predicted 
by the set NL-SH is much larger than the value adopted earlier,
where a value of $K_0= 180$ MeV has been used e.g. in BCK.

A comparison of the neutron chemical potentials $\mu_n$ with the 
results of the compressible liquid drop model suggests that the 
RMF results are not very sensitive to the value of the adopted 
compressibility coefficient. (see Sect. \ref{sec: Comp}). 

However, it is to be noted that the results presented in this work 
are meant to apply to physical conditions prevalent in the collapsing
core up to $\rho=10^{12}$ g/cm$^3$, i.e. well below the density at 
which the core reaches maximum compression and rebounds. Since the 
nuclear composition of the core at this stage can still be treated 
as consisting of individual nuclei plus Coulomb lattice corrections,
one may use a relatively high value of bulk compressibility as in NL-SH 
without worrying too much about the effects that it may have on the 
strength of the shock generated by the re-bouncing core. 

Since mean-field calculations take into account only the long-range 
part of the nucleon-nucleon interaction the short-range pairing 
correlations have to be incorporated in addition. In the present work, 
the simplistic monopole pairing interaction is used and the resulting 
equations are solved in the BCS approximation. Because the value of the
coupling constant in the pairing interaction is not known precisely,
pair-gaps obtained from odd-even mass differences are generally used as a
measure of the pairing strength. Initially, the calculations were 
started off using pair-gaps obtained by the 4-point method 
of Bohr \& Mottelson (1969) from the experimental values of 
the binding energies, wherever known. However, the experimental binding 
energies are not known for most of the nuclei of interest and we have 
adopted the pair-gaps suggested by M\"oller et al. (1995) using
isospin asymmetry arguments. For the cases where the experimental binding 
energies are known, these and the theoretically computed binding energies 
are displayed in Fig.\ref{fig3: Ch4fig1}.

In cases where multiple energy solutions were obtained, the solution 
corresponding to the lowest energy was taken as our preferred
solution. It is found that deformations of the lowest energy solution
predicted by the RMF calculations are similar to those predicted by the 
Finite Range Droplet Model (FRDM) in M\"oller et al. (1995), which leads
in this case practically to the same results as the Finite Range
Liquid Drop Model (FRLDM). 

The values of the pairing gaps $\Delta_n$ and $\Delta_p$ used in the 
entire series of calculations have been recorded in tables \ref{tab3: MnFe}, 
\ref{tab4: CoNi}, \ref{tab5: CuZn} and \ref{tab6: GaGe} for a series of 
isotopes ranging from $^{54}\rm Mn_{25}$ to $^{81}\rm Ge_{32}$. We also
show in these tables the calculated values of the quadrupole deformation 
beta $\beta$, the binding energies $BE_{RMF}$
and the chemical potentials $\mu_n$ and $\mu_p$. The binding energies
are compared with the experimental values $BE_{Exp.}$ 
The data for the Ni isotopes is especially useful because these
isotopes are spherical and in later sections we will compare the RMF 
calculations with macroscopic models as the macroscopic liquid drop model 
(BBAL, BBCW), which do not take into account effects due to deformation.

In Table \ref{tab7: NL3} we present for the chain of Ni-isotopes calculations
with the parameter set NL3 Lalazissis (1997).  For most of the elements in
this chain the binding energies obtained with the new parameter set NL3
are somewhat better, the set NL-SH however seems to be superior for
very large neutron excess. Comparing with table \ref{tab4: CoNi} we find
for the chemical potentials relatively small differences of around 
100-200 keV between the two parameter sets. We therefore do not expect 
dramatic changes in using the parameter set NL3 as compared to NL-SH, 
which is applied in all the other investigations in this work.

How was this range of isotopes chosen? The significant difference of nuclei 
in the collapsing core from those in the laboratory is that the former are 
more neutron-rich. The typical electron fraction Y$_e$ ranges from $0.42$ 
to $0.38$ in the early stages of the collapse rather than from $0.46$ to $0.5$ 
as is the case with the more symmetric laboratory nuclei. Nevertheless 
they are still far less neutron-rich than matter in neutron stars where the 
Y$_e$ could be near 0.05. The nuclear statistical equilibrium
gives rise to a fairly broad range of nuclei at any given instant of
time, and in a given mass-zone of the collapsing core which can affect the
(potentially) observable neutrino spectrum. The range of nuclei chosen here 
corresponds to values of the lepton fraction $Y_l$, which for this 
pre-trapping range of temperatures and densities, equals the electron 
fraction $Y_e$, ranging from 0.42 to 0.38. They are all mid-$f-p$ shell 
nuclei which have been shown (Hix \& Thielemann 1996)
to dominate the core nuclear configuration for this stage of the 
stellar evolution.

\section{Comparison of Chemical Potentials}
\label{sec: Comp}

If a detailed simulation of the stellar collapse is carried out, where
electron-capture and beta decay rates of individual nuclei have to be 
taken into account in a network calculation, 
the EOS based on macroscopic 
spherical liquid-drop models cannot be expected to reproduce accurately 
(i.e. with a deviation of approximately $1$ MeV) the nuclear chemical 
potentials ($\mu_n$, $\mu_p$ and $\hat \mu$) and binding energies 
(the $W_N$ values) in the low-density region. In the following sections
we investigate the amount of deviation that can be 
expected due to shell 
effects and pairing correlations on nuclear properties under stellar 
conditions. We also discuss the possible improvements which can be made 
on $W_N$ using more advanced versions like the macroscopic-microscopic
mass formulae, e.g. the Finite Range Liquid Droplet Model (FRLDM) of 
M\"oller et al. (1995), which account for deformation effects.

The RMF computations presented in Sect. \ref{sec: RMFapp} are compared 
here with values of $\mu_n$ and $\hat \mu$ calculated from the models 
discussed in the previous sections, extended to the case of isolated nuclei, 
i.e. ignoring the lattice corrections. Comparison is also made between 
RMF calculations and the finite temperature BCK-EOS, at low temperatures
and densities ($\rho_{10} = 0.1$ and $T= 0.1$ MeV) using both the values of  
$K_0$ and $S_v$ adopted by BCK in their calculations
and the more recent values used by Sharma et al. (1993).  
These comparisons show that the binding energies as well as the
neutron and proton chemical potentials have considerable differences 
from the values predicted by the macroscopic models of BBAL.
The  results from RMF calculations are in better agreement with the 
predictions from FRLDM calculations that are based on more recent 
compilations of the laboratory data on nuclei.

\subsection{Comparisons with the Spherical Liquid-Drop Models }

It has been stated earlier that the relations for $\mu_n$ and $\hat \mu$ 
in Eqs. (\ref{eq: Fullmun}) and (\ref{eq: Fullmuh}) are applicable for 
nearly symmetric, incompressible nuclear matter, confined by spherical 
surfaces. The bulk energy term per nucleon $W_{bulk}$ was taken to be 
dependent on the nuclear volume energy and on the nuclear asymmetry 
energy (see BBCW Eqs 2.1 to 2.10, especially 2.4)
given by Eq. \ref{eq: Wbulk}. A derivation of $\mu_n$ using 
this form of the bulk energy gives the following relation for the 
neutron chemical potential of a mean nucleus of mass no. $A$ :-
\begin{eqnarray}
\mu_n & = & -16 + 29.3(1-2x)^2 +117.2\,x(1-2x) \nonumber \\
&   & +W_{size}(x)
\frac{7x -3}{3(1-x)} = \mu_n|_{vol} + \mu_n|_{size},
\label{eq: BBCWmun}
\end{eqnarray}
where 
$W_{size}(x)=75.4 x^2 (1-x)^{4/3} $ 
is the 
combined surface and Coulomb energy (without lattice correction,
i.e. $f_{\rho} [u] = 1$) 
for the "mean nucleus" and 
$\rho_0(x)$ is the nuclear density. 
Fig.\ref{fig4: BBCW} displays 
the chemical potential obtained form Eq. (\ref{eq: BBCWmun}) for Ni 
isotopes, alongside with their RMF and BCK-EOS counterparts. 
It should be noted here that because of an error in the surface terms, 
the $\mu_n$ from the BBAL equations do not agree well with the RMF results 
(in fact, they are off by up to 2 MeV).

The RMF results are also compared with the BCK-EOS in Fig.\ref{fig2: ch5mn}. 
For generating this set of numbers for the BCK-EOS, the density and the 
temperature were set at 10$^9$ g/cm$^3$ and and 0.1 MeV so that 
we are as close as possible to the end-stages of Si-burning, with their 
nuclear asymmetry energy coefficient taken as $E_S= 31.5 ~\rm MeV$ 
and the incompressibility as $K=180$. However, the RMF calculations
correspond to a incompressibility coefficient $K=354.95$
and $S_v=36.1$ for reasons discussed in Sect. \ref{sec: RMFapp}.
Values for the nuclear incompressibility deduced from experiments
have shown a large variation, $K_0$ ranges from 180 to 300 MeV. 
Since the objective of the comparisons here is to evaluate the deviation
that can be expected in the LDM-EOS (compressible and incompressible) 
due to exclusion of shell and pairing effects, the $\mu_n$ and 
$\hat \mu$ from BCK-EOS were recalculated using different values of 
$S_v$ and $K_0$. It was found that the results were more sensitive 
to a variation in $S_v$ than to $K_0$. In order to obtain a suitable 
value of $S_v$ (used in BCK-EOS) which would better reproduce the 
RMF chemical potentials, a fit was made of the following expression 
of $\hat \mu|_{RMF}$:
\begin{equation}
\hat \mu= 4 S_v (1-2x) - 290 \frac{x^2 (1-x)^2}{A^{1/3}}
\left[\frac{1}{x} + \frac{2}{x}\frac{1-2x}{1-x}\right]
\label{eq: fitmuhat}
\end{equation}
This gives an "average" value of $S_v$ (over the range of isotopes for Mn, Fe, 
Co, Ni, Cu, Zn, Ga, Ge considered in the tables 
\ref{tab3: MnFe}, \ref{tab4: CoNi}, \ref{tab5: CuZn}, \ref{tab6: GaGe})  
of 30.4 MeV. It is found that using the low value of $K=180~\rm MeV$, 
the "best fitted" value of $S_v$ is 30.34 MeV. Figs.\ref{fig5: ch5bcknewn} 
and \ref{fig6: ch5bcknewh} compare $\mu_n$ and $\hat \mu$ for Ni isotopes 
using this parameter set in BCK-EOS with RMF results.

\subsection{The Finite Range Liquid Drop Model (FRLDM)}
\label{sec: FRLDM}

The various forms for the total binding energy of a 
nucleus $W_N$ that have been quoted in Sect. \ref{sec: precEOS} have all 
been for spherical nuclei without shell and pairing corrections. 
A  refinement of these simple liquid drop mass formulae of Myers \& Swiatecki 
(1966) and Myers et al. (1995) has been made using the microscopic-macroscopic 
methods. In this approach, the total nuclear potential energy is taken 
to be the sum of the energy of a term which deals with the bulk properties 
of nuclear matter and a set of terms which deal with both the 
shell and pairing effects as well as effects due to nuclear deformation. 
Here the macroscopic Finite Range Liquid Drop Model FRLDM is used to 
calculate neutron chemical potentials $\mu_n$ in the stellar core for the 
same range of nuclei for which RMF calculations were made in the previous 
section. This approach is justifiable for pre-trapping densities where 
nuclear matter is treated as an ensemble of isolated cold nuclei with a 
small drip neutron fraction. At higher densities two additional effects
would have to be taken into account: (a) the drip neutron contribution 
as given in Eq. \ref{eq: Etot} and (b) the lattice corrections made in the 
Coulomb term. The Finite Range Liquid Drop model is an extension of the 
LDM of Myers \& Swiatecki (1966) with modifications made for the finite 
range of the nuclear forces by means of a folded Yukawa potential with 
exponential term. 

Recently there has been a further modification of the FRLDM (Myers \& Swiatecki
1995) to the FRDM, the Finite Range Droplet Model (M\"oller et al. 1995). 
This has been done in order to account, among other things, 
for the corrections due to Coulomb redistribution effects and the 
effect of nuclear incompressibility on neutron and proton radii. 
This improves the predictions of higher moments of nuclear deformation 
and binding energies in the FRDM model. However, it also depends 
on a larger number of parameters. Hence, the older FRLDM model is 
used for in most of our calculations. From Fig. \ref{fig7: ch5ni}
and Fig. \ref{fig8: frdmni} we see that these models make
practically no differences for the present calculations. 

In the FRLDM model, the macroscopic component of the total nuclear 
binding energy $W_N=W_{mac}(Z,N)$ is given by
\begin{eqnarray*}
\lefteqn{W_{mac}(Z,N) = - a_v(1 - \kappa_v I^2) + a_S(1- \kappa_s I^2)B_1 A^{2/3} } \nonumber \\
& & + a_o A^0 + c_1 Z^2/  A^{1/3}B_3 - c_4 Z^{4/3}/  A^{1/3} + f(k_f r_p) Z^2/A \nonumber \\
& & -c_a (N-Z) +W \left( |I| + \left\{ \begin{array}{ll}
                    1/A &\mbox{Z=N, odd}\\
                    0   & \mbox{otherwise}
                   \end{array} \right.  \right) + \Delta_{\rm pair}, 
\end{eqnarray*}
where the average pairing energy $\Delta_{\rm pair}$ is obtained from 
nuclear binding energies using the 4-point method of Bohr et al. 
(Sect. \ref{sec: RMFapp}) and is 
given by:
\begin{equation}
\Delta_{\rm pair} = \left\{\begin{array}{ll}
\overline\Delta_p + \overline\Delta_n - \delta_{np} & \mbox{ Z odd, N odd} \\
\overline\Delta_p                                    & \mbox{ Z odd, N even} \\
\overline\Delta_n                                    &\mbox{ Z even, N odd} \\
0                                                    & \mbox {Z even, N even}
\end{array}
\right.  
\end{equation}
where $I= (N-Z)/A$ is the relative neutron-proton excess and  
$c_1=\frac{3}{5}\frac{e^2}{r_0}$ and 
$c_4=\frac{5}{4}\left(\frac{3}{2\pi}\right)^{2/3}c_1$. 
The first term $W_mac(Z, N)$
is the volume energy, the second the surface energy, 
with a correction for surface asymmetry, the third term is due to the 
Coulomb energy, the fifth is the Coulomb exchange term. The other terms 
are (in that order) due to proton form factor correction to the Coulomb 
energy, the charge asymmetry energy, and the Wigner term respectively. 
The numerical values of the various parameters used in this model are quoted 
in table \ref{tab2: FRLDM}. The proton form factor correction to the 
Coulomb energy is given by
\begin{eqnarray}
f(k_f r_p) & = & -\frac{r_p^2 e^2}{8 r_0^3} \times \nonumber \\
&  & \left[ \frac{145}{48} - \frac{327}{2880} (k_f r_p)^2 
+  \frac{1527}{1209600} (k_f r_p)^4 \right]
\end{eqnarray}
Both coefficients $B_1$ and $B_3$ are shape dependent quantities. 
The factor $B_1$ is the relative generalized surface or 
nuclear energy. It takes into account the finite range $a$ of the 
nuclear forces and is given by:
\begin{eqnarray}
\lefteqn{B_1 = \frac{A^{-\frac{2}{3}}}{8\pi^2 r_0^2 a^4} \times} \nonumber \\
& & \int\int_V 
\left(2 -\frac{|{\bf r} -{\bf r'}|}{a} \right) 
\frac{\exp(-|{\bf r} -{\bf r'}|/a}{|{\bf r} -{\bf r'}|/a}
d^3 r d^3 r'
\end{eqnarray}
and $B_3$ is the relative Coulomb energy and is given by:
\begin{eqnarray}
\lefteqn{ B_3 =\frac{15 A^{-5/3}}{32 \pi^2 r_0^5} \times } \\
& &  \int\int_V \frac{d^3 r d^3 r'}{|{\bf r}-{\bf r'}|} 
  \left[1-\left(1 + \frac{1}{2} \frac{|{\bf r} -{\bf r'}|}{a_{den}}\right)
     \exp(-|{\bf r}-{\bf r'}|/a_{den}) \right]  \nonumber
\end{eqnarray}
Using  Eq. (\ref{eq: mun}), we find the following expression for the volume
contribution to  $\mu_n$ :
\begin{equation}
 \mu_n|_{vol} = -16 + 123.04(0.5 -x)^2 + 246.08 x (0.5 -x) 
\end{equation} 
Note that the bilinear term in this expression reduces approximately 
to the linear term in Eq. (\ref{eq: Fullmun}) when symmetric nuclei are
considered ($x \simeq 0.5$). The surface etc. contributions to 
$\mu_n|_{total}$  were obtained by using Eq. (\ref{eq: mun}) on the 
surface energy, the Coulomb energy and the Coulomb exchange correction 
components of the FRLDM expression for total energy quoted above. 
The small corrections due to the shape dependent terms $B_1$ and 
$B_3$ were ignored, as were the other terms in the expression for the 
total nuclear binding energy.

In Fig.\ref{fig7: ch5ni} the functional form of $\mu_n$ in Eq. 
(\ref{eq: mun}) using FRLDM is compared with the corresponding RMF values 
for the spherical Ni isotopes and in Figs.\ref{fig2: ch5mn}, 
\ref{fig9: ch5fe}, and \ref{fig10: ch5ga} for the deformed Mn, Fe, and 
Ga isotopes. It is found that the FRLDM model reproduces the RMF values 
better than either the BCK-EOS (with the "best" parameters employed by them) 
or the spherical-LDM results in Eq. (\ref{eq: Fullmun}).
A comparison of the binding energy predictions of both the spherical, 
incompressible LDM-EOS and FRLDM model is shown in Figs.\ref{fig11: ch5nibe} 
and \ref{fig12: ch5gebe}. The latter models reproduce the binding energies 
with an accuracy of $\simeq 1$ MeV.

In Fig.\ref{fig8: frdmni} we compare the chemical potential of neutrons
in the Ni-isotopes calculated within RMF-theory using parameter NL-SH
with the results of the Finite Range Droplet Model (FRDM) of M\"oller et al.
(1995).

\section{Equation of State at Finite Temperatures}
\label{sec: FiniteTemp}
\smallskip
At low densities, the neutrons, protons and nuclei can be described 
by an ideal Boltzmann gas. The supernova matter is then in nuclear 
statistical equilibrium at fixed electron concentration Y$_e$, where 
the electrons, positrons and neutrinos (when they are trapped) are 
non-interacting Fermions (El Eid \& Hillebrandt 1980).
This approach is justified as 
long as the nucleon vapor is dilute, i.e. up to about a tenth of the 
nuclear saturation density. It does not strongly modify the nuclear 
properties. Higher densities and temperatures require an alternate 
microscopic description which accounts for interactions between all
the nucleons present. Approaches such as equilibrium in the bulk, 
the compressible liquid drop model, the Thomas-Fermi, or the 
Hartree-Fock method have been used in the literature.

In the macroscopic approach, the density of a given species
is obtained from equations of the Saha-type where one requires 
partition functions which properly account for the internal 
excitations of the nuclear species. One also needs the nuclear 
ground state energies, i.e. the binding energies of these nuclei. 
Usually one uses semi-empirical mass formulae for the binding
energies and Fermi gas based level densities for the nuclear 
partition functions. The quality of these ingredients determine 
the level of accuracy of the EOS and its range of applicability.
While the macroscopic method may be adequate for the computation 
of stellar collapse dynamics, one has to take into account shell-effects
and pairing correlations in order to calculate threshold Q-values for 
electron captures on nuclei, if one is interested in the determination
of the neutrino spectrum. In addition, ideally one would like to have 
an approach which is applicable in both low and high density situations 
encountered in supernova physics. From this point of view a microscopic 
approach like the RMF method, which satisfies both criteria, is a 
preferred method. It can also be used to compare results obtained
from other microscopic methods. 

It can be shown that in the case of low entropy of the stellar core 
matter, e.g. at the stage of collapse, a multi-phase equilibrium
such as between nuclei and nucleonic vapor can be conveniently 
studied by using a low temperature expansion for the dense nuclei, 
along with a high temperature expansion for the nucleonic vapor 
(Vautherin 1994). Vautherin's description of the problem 
is a mean-field approximation and a simple extension of our RMF 
calculations to finite temperatures is thus possible.
  
For a spin saturated, symmetric nucleus (N=Z) the mean field potential 
from a Skyrme force with a zero range two-body attraction and a 
three-body repulsion, with coefficients $t_0 = -983.4$ MeV fm$^3$ 
and $t_3 = 13105.8$ MeV fm$^6$ is given by (Ring \& Schuck 1980):
\begin{equation}
U({\bf r}) = \frac{3}{4} t_0 \rho({\bf r}) + 
\frac{3}{16} t_3 \rho^2 ({\bf r}). 
\label{eq: Ur}
\end{equation}
With this potential the single particle orbitals $\phi_i({\bf r}$) 
are given by the solution of the
Schr\"odinger equation with eigenvalues $e_i$. The Fermi occupation 
numbers $f_i$ at finite temperatures are given by
$f_i = (1 + \exp ((e_i - \mu)/kT))^{-1}$ and the chemical potential 
$\mu$ is fixed from the particle number $ \sum_i f_i = A $. 
This gives the self consistent nucleon density at a finite temperature
as $\rho({\bf r}) = \sum_i f_i \left| \phi_i ({\bf r}) \right|^2$.
The entropy of the nucleus obtained in a similar way by the expression 
for independent fermions:
\begin{equation}
S ~=~ - k \sum_i \{ f_i \log f_i + (1 - f_i) \log(1 -f_i) \}.
\end{equation}
For the uniform nuclear matter approximation, the density of the nucleons 
and kinetic energy densities are found as
\begin{equation}
\rho = g \lambda^{-3} F_{3/2}(\beta U - \alpha) 
\end{equation}
and
\begin{equation}
\tau = 4 \pi g \lambda^{-5} F_{5/2} (\beta U - \alpha) 
\end{equation} 
where g=4 is the spin and isospin degeneracy factor $(2S+1)(2I+1)$, 
$\lambda = (2\pi \hbar^2/m kT)^{1/2})$ is the thermal wave length, 
$\beta =1/kT$, and $\alpha = \beta \mu$. The Fermi-integral $F_n$ 
is given by:
\begin{equation}
F_n (x) = (2/\sqrt\pi) \int_0^{\infty} y^{n-1} (1 + \exp(y+x))^{-1} dy. 
\end{equation}
Once the density and the temperature of stellar matter is specified, the
argument $(\beta U - \alpha)$ of the Fermi function is determined, which
in turn can be used to compute kinetic energy density, entropy density,
pressure, and the free energy density of the system.

A consideration of the phase diagram, i.e. the isothermal lines in 
$P-\rho$ plane, indicates that at temperatures below a critical temperature
a configuration with constant density does not correspond to the lowest 
value of the free energy for some density ranges. One has a two phase 
system consisting of a uniform density nucleus and the nucleonic vapor 
in equilibrium with each other as a more favorable state. Under such 
conditions, one has:
\begin{equation}
\mu(\rho_N, T) = \mu(\rho_v, T)~~~~~{\rm and}~~~~P(\rho_N, T) = P(\rho_v, T).  
\end{equation}
So far, two effects which are important for nuclei have been left out, 
the finite size (or surface) effects and the Coulomb interactions.
Levit and collaborators (Levit \& Bonche 1985; Besprosvany \& Levit 1989)
have considered
the nucleus as a homogeneous system of $A = N + Z$ nucleons inside a
spherical volume $\Omega$ of radius R which is in equilibrium with the
nucleonic vapor of volume $\Omega_v$ and density $\rho_v$. 
Because of the large incompressibility of nuclear matter it can be shown 
that the change in the density of the nuclear phase caused by Coulomb 
and surface tension effects is rather small. Therefore in the 
equilibrium equations between the nuclear and vapor phases, 
it is adequate to evaluate the Coulomb and surface terms for the 
pressure and the chemical potentials at saturation density $\rho_0$ and 
$R=R_0=(3/4\pi \rho_0)^{1/3}$ (Vautherin 1994). Using the Gibbs-Duhem relation 
$\rho \partial \mu / \partial \rho = \partial P / \partial \rho$ 
one can also eliminate the density of the nucleus phase in the vapor-nucleus
equilibrium equations, and observing that the coefficients of 
$\rho- \rho_0$ 
in the 
expansion of the nuclear pressure and of the chemical potential 
have to be equal. Thus one has:
\begin{equation}
\mu(\rho_v, T)~=~ \frac{E_0}{A} - \frac{\pi^2}{4} \frac{(kT)^2}{T_F} 
+ \frac{1}{R_0}\left(\frac{2 \alpha(T)}{\rho_0} + \frac{Z^2 e^2}{A}\right) 
\end{equation} 

The RMF calculations outlined in the previous sections automatically account
for the surface and Coulomb effects in the nuclear chemical potential
calculations (such as the last term within parentheses in the above
equation). Since $T_c$ is quite high, it is adequate to use in the
computation of the equation of state at low densities the zero temperature 
surface tension effects already built into the RMF code for the nuclei, 
i.e. $\alpha(T) =\alpha(0)$. Therefore the only finite temperature
correction is the second term on the RHS of the above equation.

\subsection{Lattice Size Effects} 

As an  attempt to extend the RMF results presented in section
\ref{sec: RMFapp}, to higher densities, the density dependent lattice 
correction should be incorporated into the Laplace equation for the 
electric potential:
\begin{equation}
-\Delta A^0 ( {\bf r}) = \rho_p ({\bf r}) - \rho_e ({\bf r})
\end{equation}
where $A^0$ is the Coulomb potential, $\rho_p ({\bf r})$ is the proton
charge density and $\rho_e ({\bf r})$ is the electron charge distribution
density in the Wigner-Seitz cell. Here, the lattice correction has been 
incorporated into the Coulomb energy part of the total energy
by the inclusion of the multiplicative factor $F_{lattice}= (1-\frac{3}{2}
(\frac{\rho}{\rho_s})^{1/3}+\frac{1}{2} (\frac{\rho}{\rho_s}) )$ where $\rho$
and $\rho_s$ are the stellar matter and nuclear saturation densities 
respectively. In Fig.\ref{fig13: ch5nimuhlat} we show the values of 
$\hat \mu$ for Ni isotopes along with the results from Eqs. 
(\ref{eq: Fullmun}) and (\ref{eq: Fullmuh}) for the matter density 
$\rho$ = $10^{12} ~\rm g cm^{-3}$.

At densities near $\approx 10^{12}$ g cm$^{-3}$ the distance between
individual nuclei becomes so short that the nuclear motions begin to
be correlated due to the Coulomb interaction between the nuclei and 
at somewhat higher densities the nuclei arrange themselves on the lattice 
sites of a crystal. Thus, towards the end of the density regime that 
one would be interested in for computing early (un-scattered) neutrino 
spectra, EOS calculations have to take into account screened Coulomb 
interactions. This is usually done by calculating thermodynamic 
quantities in the Wigner-Seitz approximation. Stellar matter with
a plasma parameter $\Gamma = Z^2 e^2 /R kT \geq 155$, arranges itself 
in a lattice and the Coulomb interaction energy can be written at zero 
temperature as:
\begin{equation}
 E_c =\frac{3}{5}\frac{Z^2 e^2}{R_N} \left\{ 1-\frac{3R_N}{2R_c} + 
\frac{1}{2}\left(\frac{R_N}{R_c}\right)^3 \right\} 
\end{equation}
This therefore motivates the inclusion of the density dependent
$F_{lattice}$ correction in the RMF equations. 

\subsection{Asymmetric Nuclear Matter and 
Approximate Solution of the Equilibrium Equations}

To describe nuclei with different neutron and proton numbers as in the
neutron-rich situation encountered in the stellar collapse one has to 
include the bulk asymmetry term $a_{\tau}x_N^2$ in the mass formula, 
where $x_N=(N-Z)/A$ is the nuclear asymmetry parameter. By writing the 
equilibrium equations explicitly in terms of low and high temperature 
expansions in the nuclear and in the vapor nucleonic chemical potentials, 
one is able to relate the difference in the nuclear component of the 
neutron and proton chemical potentials to the asymmetry parameter $x_N$ as:
\begin{equation}
\mu_{Nn} - \mu_{Np} = 4 a_{\tau} (T) x_N,  \label{eq: V3.17}
\end{equation}
where, $a_{\tau}(T)$ is the second derivative of the free energy with 
respect to $x_N$:
\begin{equation}
a_{\tau} (T) = \frac{1}{3} T_N - \frac{1}{4} t_0 (x_0 + 1/2)\rho_N -
\frac{1}{16} t_3 \rho^2_N 
+ \frac{\pi^2}{36} \frac{(kT)^2}{T_N}  
\label{eq: V3.18} 
\end{equation}
where $x_0 = 0.48$.
Similarly the vapor asymmetry $x_v$ and  nuclear asymmetry 
can be related as:
\begin{equation}
x_v = \tanh (2a_{\tau} x_N /kT), 
\end{equation}
while the vapor density can be related to the average nuclear 
chemical potential $\mu_N =(\mu_{Nn} + \mu_{Np})/2$:
\begin{equation}
 \rho_v = g \lambda^{-3} (1 - x_v^2)^{-1/2} exp(\mu_N/kT). 
\end{equation}
Here $\mu_N$ defined above can be approximated to the
first order in $x_N$ as:
\begin{equation}
\mu_N = 
\frac{E_0}{A} + \frac{1}{9} \frac{\rho_N - \rho_0}{\rho_0} 
\left( K - \frac{5\pi^2(kT)^2}{2 T_F}\right) - \frac{\pi^2(kT)^2}{12 T_F}.
\end{equation}
It has been shown (Vautherin 1994) that it is a good approximation to
replace $a_{\tau}$ by its value at zero temperature and normal density, 
since the dependence of $a_{\tau}$ on temperature is weak. The pressure 
in the vapor phase is given by the high temperature approximation:
\begin{eqnarray}
P_v  & = & \rho_v T \left( 1 + \frac{\rho_v \lambda^3}{4 g \sqrt 2} 
(1 + x_v^2) + \cdots
\right) \nonumber \\
& & + \frac{1}{8} t_0 \rho^2_v (3 - x_v^2 (2 x_0 +1)) 
+ \frac{1}{8} t_3 \rho^3_v (1 - x_v^2).
\end{eqnarray}
Although the total pressure is dominated by the lepton pressure,
one may estimate the nucleonic component from the above expression.

\section{Summary}
\label{sec: Summ}

Using the relativistic mean field method, we have evaluated in this paper 
properties of individual nuclear species which play a significant role 
for the physical and thermodynamic properties as well as the dynamics of 
a pre-supernova collapsing stellar core. In particular, we have determined 
the nuclear chemical potentials $\mu_n$ and $\mu_p$, which, along with 
the nuclear weak interaction strengths control the neutronization of the 
matter. We have focussed our interest on properties of nuclei 
present in the core at a density range $10^9$ to $10^{12}$ g/cm$^3$ which 
corresponds to the temperature range of $\simeq 0.2$ to $1.5 $ MeV. In
this range the core composition is dominated by heavy, neutron-rich, 
$f-p$ shell nuclei in a Saha equilibrium, with a low density vapor of 
free neutrons, and an even lower density vapor of free protons. In 
this temperature and density range, matter can be treated as an ensemble 
of non-interacting Boltzmann particles, at least at $\rho_{10}=0.1 \ - \ 100$
$g/cm^3$. This implies that nuclear properties under these conditions are 
determined by those of individual nuclei, and structure effects like shell 
effects, pairing and deformation play a significant role. 

Our study has been motivated by the fact that the existing equations of
state used for numerical studies of stellar core collapse are based on 
macroscopic liquid-drop models, which generally do not take into account 
shell, pairing and other nuclear structure effects - which play an important 
role at lower energies. These EOS are suitable for matter at $T \gg 1.5$ MeV, 
and $\rho \gg 10^{12}$ g/cm$^3$, when the nuclear matter undergoes a phase 
transition from the state where individual nuclei are present in a gas of 
dripped nucleons, to the state where the nuclei merge into infinite nuclear 
matter (see Sect. \ref{sec: Intro} and (BBAL, BBCW, Cooperstein (1985)).
Because shell and pairing effects are already washed out at temperatures 
beyond $1$ MeV and at densities beyond trapping density i.e. 
$10^{12}$ g/cm$^3$, ideally, a low-density, zero temperature EOS would 
be desirable, which smoothly goes over to the high density, high 
temperature equations of state based on the liquid drop model 
(see Swesty et al. 1994).

In order to account for the shell and pairing effects persistent 
at low temperatures we studied the nuclear structure effects in the 
framework of relativistic mean field theory, using a Lagrangian with
non-linear self-interactions for the $\sigma$-meson. This method has been 
shown to provide a good description of the ground state properties of nuclei 
(Gambhir et al. 1990). The nuclei of interest range from $^{54}$Mn$_{25}$
to $^{81}$Ge$_{32}$ (see Tables \ref{tab3: MnFe}, \ref{tab4: CoNi}, 
\ref{tab5: CuZn}, and \ref{tab6: GaGe}), predominantly neutron-rich 
$f-p$ shell nuclei. It is the first time that the RMF method is used 
to calculate the binding energies of nuclei with a neutron excess
in the range of interest. 

In order to test the validity of high-density equations of state at 
low densities BBAL we compare the chemical potentials 
$\mu_n$ and $\hat \mu$ with those obtained in RMF theory and we
find considerable deviations. To study the influence of nuclear
compressibility we have also compared  the chemical potentials 
$\mu_n|_{RMF}$ and $\mu_p|_{RMF}$ with those of the compressible 
liquid-drop model based on the EOS of BCK. To test the 
effects of a variation in the nuclear incompressibility $K_0$ and in 
the volume symmetry energy $S_v$, we computed the results of  
BCK with a range of values of $K_0$ and $S_v$. We find that 
the results are more sensitive to the variation in $S_v$ than in 
$K_0$. From a fit of $\hat\mu|_{RMF}$ to the analytical expression 
for $\hat\mu|_{BCK}$, we find an "averaged" value of $S_v$ of 30.34. 
Further, to account for nuclear deformation effects we use the FRLDM 
expression for the nuclear binding energy to compute $\mu_n$ and 
$\hat \mu$. A comparison of RMF values with this "extended" macroscopic 
model demonstrates the influence of shell and pairing effects on 
$\mu_n$ and on $\hat \mu$ (see Figs.\ref{fig11: ch5nibe} 
and \ref{fig12: ch5gebe}). 

Finally, in order to take into account the
first order Coulomb correlations 
between the nuclei at high densities, we include a lattice correction in 
our RMF computations of $\hat\mu$. 
A comparison of these lattice corrected RMF results 
with the BBAL/Fuller EOS for $\hat\mu$
is shown in Fig.\ref{fig13: ch5nimuhlat}.
We have also outlined the reasons for obtaining the relatively small
corrections that are needed for finite temperature corrections to the
EOS at low entropies relevant for the early stages of a massive star's
core collapse. Methods to generalize the results to finite temperature
EOS calculations, using the chemical potentials and other quantities 
obtained from RMF calculations have been indicated. 


Many groups have
computed and analysed the collapse of the stellar core of a massive star 
to evaluate the effects of various physical processes in determining
conditions prior to core-bounce. 
As collapse is shown to be a nearly homologous one (see
BBAL and references therein), a one-zone collapse is adequate
for the investigation of thermodynamic and nuclear properties
(see e.g. EH (1980); Hillebrandt et al. (1984) which used the EOS of Wolff
\& Hillebrandt (1983); Ray et al. (1984); Fuller (1982);
etc). The EOS of dense stellar matter
affects the electron capture rate and the evolution of the lepton fraction
mainly through the influence of the chemical potentials of the free
nucleons (see Murphy (1980); Sutaria \& Ray (1997)). 
For example, the thermodynamic quantities
at the beginning of stellar collapse for the RMF results
and those of EH are compared in Table \ref{tab8: RMF-EH}.
However, the intrinsic electron capture rates on nuclei
employed by different authors have been different and have evolved
considerably over the years. It is therefore difficult to extract
the purely EOS related effects on neutronisation of different
calculations.
The stellar collapse calculation using the full RMF EOS in a self-consistent
manner with the latest electron capture rates
is in preparation and the details of    
the evolution of nuclear and thermodynamic variables will be published 
elsewhere.
 
\begin{acknowledgements}
This research is part of the 9th Five Year Plan project 9P-208[a] 
at Tata Institute of Fundamental Research. AR was a
U.S. National Research Council Senior Research Associate during
part of the time when this research was carried out. 
Part of the work has been supported by the Bundesministerium
f\"ur Bildung und Forschung under the contract 06 TM 875.
\end{acknowledgements}



\begin{table}[t]
\begin{center}
\caption{The nuclear charge radii for Ni isotopes. The table gives the 
experimental values and those computed in this work from RMF calculations 
with the parameter sets NL1 and NL-SH.
\label{tab1: ch-radii} }
\vskip 1 true cm
\begin{tabular}{cccc}
Neutron & Experimental  & RMF   & RMF  \\
number& Nadjakov et al. (1994) & (NL1) & (NL-SH) \\
\hline
30 & 3.7827 $\pm$ 0.0036& 3.804 & 3.791\\
32 & 3.8177 $\pm$ 0.0047& 3.797 & 3.794\\
33 & 3.8283 $\pm$ 0.0030& 3.805 & 3.802\\
34 & 3.8475 $\pm$ 0.0050& 3.812 & 3.809\\
36 & 3.8678 $\pm$ 0.0051& 3.830 & 3.826\\
\hline
\end{tabular}
\end{center}
\end{table}

\begin{table}[t]
\begin{center}
\caption{ The parameter set for macroscopic FRLDM model 
\label{tab2: FRLDM} }
\vskip 1 true cm
\begin{tabular}{clcc}
\hline
Quantity & Brief definition & Value (Unit)\\
\hline
$a_v$      & Volume energy constant. & 16.00126  MeV \\
$\kappa_v$      &Volume asymmetry constant. & 1.9224  MeV\\
$a_s$      &Surface energy constant. & 21.18466  MeV\\
$\kappa_s$      &Surface asymmetry constant. & 2.345  MeV\\
$a_0$      & $A^0$ constant.             & 2.615  MeV\\
$c_a$      & Charge asymmetry constant. &0.10289  MeV\\
$r_0$        & Nuclear-radius constant.  & 1.16  fm\\
$r_p$        & Proton r.m.s. radius.     &0.80 fm\\
$a$          & Range of                & 0.68 fm \\
             & Yukawa-plus-exponential &   \\
             & potential.              &   \\
$a_{den}$   & Range of Yukawa  
& 0.70 fm\\
  &  function for generating &  \\
&  charge distribution. &  \\
$e^2$       & square of the e$^-$-charge & 1.4399764  MeV fm \\
\hline
\end{tabular}
\end{center}
\end{table}
\clearpage

\begin{table}[h]
\begin{center}
\caption{
Results of RMF calculations for various isotopes of Mn and Fe, giving 
the values of $\mu_n$, $\mu_p$ and the deformation parameter $\beta$.  
Also tabulated are the values of the fixed neutron and proton pairing-gaps 
($\Delta_n$ and $\Delta_p$) used in these calculations. The last two columns 
display the experimental and calculated values of the binding energies 
($BE_{\rm Exp.}$ and $BE_{\rm RMF}$ 
respectively)  
\label{tab3: MnFe}}
\begin{tabular}{cccccccc}
Isotope & $\Delta_n$  & $\Delta_p$ & $\mu_n$ & $\mu_p$ & $\beta$ & $BE_{\rm 
Exp.}$ & 
$BE_{\rm RMF}$ \\
\hline
$^{54}{\rm Mn}_{25}$&  1.00&  1.50&  -8.6790&  -8.6260&  0.1265  &-471.84940  
&-471.64801\\
$^{55}{\rm Mn}_{25}$&  1.00&  1.50&  -8.7800&  -9.2290&  0.2082  &-482.07599  
&-480.91101\\
$^{56}{\rm Mn}_{25}$&  1.20&  1.50&  -7.9420&  -9.9240&  0.2187  &-489.34659  
&-488.92999\\
$^{57}{\rm Mn}_{25}$&  2.00&  1.50&  -7.9260& -10.5230&  0.2091  &-497.99701  
&-497.51801\\
$^{58}{\rm Mn}_{25}$&  1.20&  1.50&  -7.3470& -11.4300&  0.2213  &-504.41000  
&-503.65399\\
$^{59}{\rm Mn}_{25}$&  2.00&  1.25&  -7.1270& -11.9500&  0.2130  &-512.12903  
&-511.54700\\
$^{60}{\rm Mn}_{25}$&  1.20&  1.25&  -6.2840& -12.8120&  0.2110  &-517.62000  
&-516.78497\\
$^{61}{\rm Mn}_{25}$&  2.00&  1.25&  -6.3820& -13.1600&  0.1896  & --         
&-524.057  \\
$^{62}{\rm Mn}_{25}$&  1.25&  1.25&  -6.0970& -13.7840&  0.1354  & --         
&-528.286  \\
$^{63}{\rm Mn}_{25}$&  1.50&  1.00&  -6.0110& -14.3490&  0.1097  & --         
&-534.347  \\
$^{64}{\rm Mn}_{25}$&  1.00&  0.50&  -5.8150& -15.0530&  0.0830  & --         
&-539.142  \\
$^{65}{\rm Mn}_{25}$&  0.75&  0.50&  -5.2780& -15.5570&  0.0458  & --         
&-544.673  \\
\hline
$^{54}{\rm Fe}_{26}$&  2.50&  1.50& -12.0530&  -7.1700&  0.0003  &-471.76392  
&-470.85400\\
$^{55}{\rm Fe}_{26}$&  1.50&  2.50& -10.5820&  -7.8950&  0.0000  &-481.06192  
&-481.15302\\
$^{56}{\rm Fe}_{26}$&  2.50&  2.50&  -9.5510&  -8.2400&  0.0000  &-492.25992  
&-492.17700\\
$^{57}{\rm Fe}_{26}$&  2.00&  2.50&  -8.9800&  -8.7980&  0.1531  &-499.90610  
&-500.02802\\
$^{58}{\rm Fe}_{26}$&  2.50&  2.50&  -8.7690&  -9.3750&  0.1514  &-509.95081  
&-509.32599\\
$^{59}{\rm Fe}_{26}$&  2.00&  2.00&  -8.1790& -10.0260&  0.1915  &-516.53180  
&-515.96600\\
$^{60}{\rm Fe}_{26}$&  2.25&  2.00&  -7.8740& -10.6300&  0.1839  &-525.34802  
&-524.05402\\
$^{61}{\rm Fe}_{26}$&  1.75&  2.00&  -7.2590& -11.3220&  0.1853  &-530.93201  
&-530.29102\\
$^{62}{\rm Fe}_{26}$&  2.00&  2.00&  -7.1360& -11.9780&  0.1421  &-538.97998  
&-537.29602\\
$^{63}{\rm Fe}_{26}$&  1.75&  1.75&  -7.0720& -13.0450& -0.0938  &-543.34998  
&-542.30103\\
\hline
\end{tabular}
\end{center}
\end{table}
\clearpage

\begin{table}[t]
\begin{center}
\caption{ 
Results of RMF calculations for various isotopes of Co and Ni, 
giving the values of $\mu_n$, $\mu_p$ and the deformation parameter
$\beta$. Also tabulated are the values of the fixed neutron and proton 
pairing-gaps ($\Delta_n$ and $\Delta_p$) used in these calculations. 
The last two columns show the calculated and experimental values of 
binding energies.   
\label{tab4: CoNi}}
\begin{tabular}{cccccccc}
Isotope & $\Delta_n$  & $\Delta_p$ & $\mu_n$ & $\mu_p$ & $\beta$ &$ BE_{\rm 
RMF}$ 
& $ BE_{\rm Exp.}$ \\
\hline
$^{56}{\rm Co}_{27}$&  1.00&  1.50&  -9.2070&  -7.1380&  0.0003  &-486.91150  
&-487.96399\\
$^{57}{\rm Co}_{27}$&  1.02&  1.50&  -9.6860&  -7.2640&  0.1408  &-498.28741  
&-497.01001\\
$^{58}{\rm Co}_{27}$&  1.17&  1.50&  -9.2300&  -7.7340&  0.1675  &-506.86050  
&-506.35599\\
$^{59}{\rm Co}_{27}$&  2.00&  1.50&  -9.1760&  -8.4530&  0.1430  &-517.31409  
&-516.29102\\
$^{60}{\rm Co}_{27}$&  1.20&  1.25&  -8.6380&  -8.8190&  0.1820  &-524.80615  
&-523.69702\\
$^{61}{\rm Co}_{27}$&  2.00&  1.25&  -8.4150&  -9.6480&  0.1520  &-534..12738  
&-532.99597\\
$^{62}{\rm Co}_{27}$&  1.15&  1.25&  -7.6400& -10.1580&  0.1640  &-540.72498  
&-539.52698\\
$^{63}{\rm Co}_{27}$&  2.00&  1.25&  -7.9050& -11.4190&  0.0915  &-549.21198  
&-548.28497\\
$^{64}{\rm Co}_{27}$&  1.20&  1.25&  -7.8090& -12.4110&  0.0455  &-555.23602  
&-554.36798\\
$^{65}{\rm Co}_{27}$&  1.50&  1.00&  -7.4920& -13.2190&  0.0011  &-562.67999  
&-562.22900\\
\hline
$^{58}{\rm Ni}_{28}$&  2.50&  2.50& -10.8710&  -5.7110&  0.0000  &-506.45901  
&-507.02899\\
$^{59}{\rm Ni}_{28}$&  2.00&  2.00&  -9.9250&  -6.3600&  0.0000  &-515.45892  
&-514.91101\\
$^{60}{\rm Ni}_{28}$&  2.50&  2.00&  -9.8810&  -6.9550&  0.0000  &-526.84735  
&-525.61603\\
$^{61}{\rm Ni}_{28}$&  2.00&  2.00&  -9.3950&  -7.6600&  0.0000  &-534.66760  
&-533.20502\\
$^{62}{\rm Ni}_{28}$&  2.25&  1.75&  -9.2210&  -8.2490&  0.0000  &-545.26501  
&-542.88800\\
$^{63}{\rm Ni}_{28}$&  1.75&  1.50&  -8.9080&  -8.9340&  0.0000  &-552.10352  
&-550.31299\\
$^{64}{\rm Ni}_{28}$&  2.00&  1.50&  -8.6650&  -9.4930&  0.0000  &-561.76013  
&-559.63000\\
$^{65}{\rm Ni}_{28}$&  1.75&  1.50&  -8.3620& -10.1220&  0.0000  &-567.85834  
&-567.52698\\
$^{66}{\rm Ni}_{28}$&  1.00&  1.00&  -8.0750& -10.7160&  0.0000  &-576.83301  
&-574.58398\\
$^{67}{\rm Ni}_{28}$&  1.50&  1.00&  -7.5840& -11.2130&  0.0000  &-582.61902  
&-583.01501\\
$^{68}{\rm Ni}_{28}$&  0.75&  1.00&  -6.9360& -11.7330&  0.0000  &-590.42999  
&-589.70001\\
\hline
\end{tabular}
\end{center}
\end{table}
\clearpage

\voffset 0.2 true in
\begin{table}[t]
\begin{center}
\caption{
Results of RMF calculations for various isotopes of Cu and Zn, 
giving the values of $\mu_n$, $\mu_p$ and the deformation parameter 
$\beta$. Also tabulated are the values of the fixed neutron and proton
pairing-gaps ($\Delta_n$ and $\Delta_p$) used in these calculations, 
as well as the experimental and calculated values of the binding energies. 
\label{tab5: CuZn}}    
\begin{tabular}{cccccccc}
Isotope & $\Delta_n$  & $\Delta_p$ & $\mu_n$ & $\mu_p$ & $\beta$ &$ BE_{\rm 
Exp.}$ 
& $ BE_{\rm RMF} $  \\
\hline  
$^{57}{\rm Cu}_{29}$&  1.00&  1.50& -14.0190&  -1.3950& -0.0009  &-484.76001  
&-485.70700\\
$^{58}{\rm Cu}_{29}$&  2.00&  1.75& -11.0700&  -2.5080&  0.0037  &-497.11401  
&-497.35001\\
$^{59}{\rm Cu}_{29}$&  2.00&  1.75& -11.2610&  -4.0470&  0.1461  &-509.87601  
&-508.19299\\
$^{60}{\rm Cu}_{29}$&  2.00&  1.75& -10.6050&  -4.9190&  0.1650  &-519.93811  
&-518.96899\\
$^{61}{\rm Cu}_{29}$&  2.25&  1.75& -10.5020&  -5.3060&  0.1394  &-531.64679  
&-530.08899\\
$^{62}{\rm Cu}_{29}$&  1.75&  1.75&  -9.8260&  -6.3300&  0.1750  &-540.53400  
&-538.87299\\
$^{63}{\rm Cu}_{29}$&  2.50&  1.50&  -9.6850&  -6.4610&  0.1374  &-551.38715  
&-549.16901\\
$^{64}{\rm Cu}_{29}$&  2.00&  1.50&  -9.0460&  -7.2040&  0.1424  &-559.30316  
&-556.85797\\
$^{65}{\rm Cu}_{29}$&  2.25&  1.50&  -9.1590&  -7.3710& -0.1001  &-569.21216  
&-567.38501\\
$^{66}{\rm Cu}_{29}$&  1.50&  1.50&  -8.8200&  -8.0180& -0.0922  &-576.27814  
&-574.91803\\
$^{67}{\rm Cu}_{29}$&  2.25&  1.25&  -8.5780&  -7.9580& -0.0083  &-585.39697  
&-583.46503\\
\hline
$^{64}{\rm Zn}_{30}$&  2.50&  2.50& -10.6040&  -5.8830&  0.1700  &-559.09912  
&-557.58899\\
$^{65}{\rm Zn}_{30}$&  2.00&  2.50&  -9.8550&  -6.7130& -0.1663  &-567.07898  
&-566.32001\\
$^{66}{\rm Zn}_{30}$&  2.50&  2.50&  -9.7730&  -7.0710& -0.1314  &-578.13885  
&-576.87000\\
$^{67}{\rm Zn}_{30}$&  2.00&  2.50&  -9.4760&  -7.5290& -0.1161  &-585.19086  
&-584.99200\\
$^{68}{\rm Zn}_{30}$&  2.25&  2.25&  -9.2010&  -7.8250& -0.0793  &-595.38898  
&-593.99298\\
$^{69}{\rm Zn}_{30}$&  1.75&  2.00&  -8.7400&  -8.2670&  0.0159  &-601.87122  
&-601.14203\\
$^{70}{\rm Zn}_{30}$&  2.25&  2.00&  -8.4010&  -8.8290&  0.0169  &-611.08600  
&-610.53101\\
$^{71}{\rm Zn}_{30}$&  1.75&  2.00&  -7.8710&  -9.4300&  0.0090  &-616.91901  
&-617.31897\\
$^{72}{\rm Zn}_{30}$&  2.00&  1.50&  -7.6870&  -9.9300&  0.0764  &-625.80200  
&-624.49799\\
$^{73}{\rm Zn}_{30}$&  1.25&  1.50&  -7.6010& -11.0730&  0.0000  &-631.15002  
&-630.94098\\
$^{74}{\rm Zn}_{30}$&  1.50&  1.50&  -7.2610& -11.7630&  0.0000  &-639.51898  
&-638.95001\\
$^{75}{\rm Zn}_{30}$&  1.25&  1.50&  -6.9250& -12.4690&  0.0000  &-644.58002  
&-645.87000\\
$^{76}{\rm Zn}_{30}$&  1.00&  1.00&  -6.4360& -13.0960&  0.0001  &-652.41998  
&-652.28497\\
$^{77}{\rm Zn}_{30}$&  0.20&  0.50&  -5.8320& -13.7630&  0.1720  &-656.94000  
&-657.80603\\
$^{78}{\rm Zn}_{30}$&  0.50&  0.50&  -5.4660& -14.2850&  0.1497  &-664.06000  
&-663.66101\\
\hline
\end{tabular}
\end{center}
\end{table}
\clearpage

\begin{table}[t]
\begin{center}
\caption{
Results of RMF calculations for various isotopes of Ga and Ge, 
giving the values of $\mu_n$, $\mu_p$ and the deformation parameter
$\beta$. Also tabulated are the values of the fixed neutron and proton
pairing-gaps ($\Delta_n$ and $\Delta_p$) used in these calculations 
as well as the calculated and the experimental values of the binding 
energies, $BE_{\rm Exp.}$ and $BE_{\rm RMF}$ respectively.    
\label{tab6: GaGe}}
\begin{tabular}{cccccccc}
\hline
Isotope & $\Delta_n$  & $\Delta_p$ & $\mu_n$ & $\mu_p$ & $\beta$ &$BE_{\rm 
Exp.}$ & 
$BE_{\rm RMF}$ \\
\hline
$^{69}{\rm Ga}_{31}$&  2.50&  2.50&  -9.7100&  -7.2800&  0.0897  &-601.99402  
&-602.03601\\
$^{70}{\rm Ga}_{31}$&  1.50&  2.50&  -9.2860&  -7.7540&  0.0010  &-609.64899  
&-609.17798\\
$^{71}{\rm Ga}_{31}$&  2.50&  2.50&  -9.0280&  -8.3030&  0.0148  &-618.95618  
&-620.22803\\
$^{72}{\rm Ga}_{31}$&  1.50&  2.50&  -8.3770&  -8.8990&  0.0014  &-625.47729  
&-626.45203\\
$^{73}{\rm Ga}_{31}$&  2.50&  1.50&  -8.4390&  -9.4400&  0.1686  &-634.66302  
&-635.96198\\
$^{74}{\rm Ga}_{31}$&  1.50&  1.50&  -8.1890& -10.0830&  0.1770  &-641.08002  
&-641.78699\\
$^{75}{\rm Ga}_{31}$&  2.00&  1.30&  -7.8210& -10.5600&  0.1730  &-649.56598  
&-650.75098\\
$^{76}{\rm Ga}_{31}$&  1.00&  1.10&  -7.6520& -11.0730&  0.1800  &-655.62000  
&-656.83698\\
$^{77}{\rm Ga}_{31}$&  0.75&  0.75&  -7.1730& -11.6270&  0.1790  &-663.65997  
&-664.02197\\
$^{78}{\rm Ga}_{31}$&  0.50&  0.50&  -6.7420& -12.4590&  0.1680  &-668.88000  
&-670.63800\\
$^{79}{\rm Ga}_{31}$&  0.25&  0.25&  -6.0880& -13.3830&  0.1540  &-676.14001  
&-677.14899\\
$^{80}{\rm Ga}_{31}$&  0.20&  0.20&  -6.9860& -13.4700& -0.0771  &-680.84003  
&-681.55902\\
$^{81}{\rm Ga}_{31}$&  0.20&  0.20&  -5.0490& -14.1430& -0.0574  &-687.51001  
&-688.55798\\
\hline
$^{70}{\rm Ge}_{32}$&  2.50&  2.50& -10.1990&  -6.7730& -0.1831  &-610.54401  
&-609.46198\\
$^{71}{\rm Ge}_{32}$&  1.00&  2.50&  -9.7040&  -7.4030& -0.1692  &-617.96600  
&-616.29401\\
$^{72}{\rm Ge}_{32}$&  2.50&  2.50&  -9.4920&  -7.8830& -0.1650  &-628.70801  
&-628.55902\\
$^{73}{\rm Ge}_{32}$&  1.00&  2.50&  -8.8660&  -8.6950& -0.1935  &-635.50000  
&-635.09198\\
$^{74}{\rm Ge}_{32}$&  2.00&  2.00&  -8.6880&  -8.9680& -0.1749  &-645.69202  
&-644.26599\\
$^{75}{\rm Ge}_{32}$&  1.00&  2.00&  -8.8110&  -9.3880&  0.1881  &-652.20398  
&-650.90198\\
$^{76}{\rm Ge}_{32}$&  1.50&  2.00&  -8.4500&  -9.9400&  0.1721  &-661.62598  
&-660.47302\\
$^{77}{\rm Ge}_{32}$&  1.00&  1.50&  -8.2590& -10.3350&  0.1793  &-667.69800  
&-667.63300\\
$^{78}{\rm Ge}_{32}$&  0.75&  1.00&  -7.9020& -10.8030&  0.1760  &-676.41998  
&-675.31500\\
$^{79}{\rm Ge}_{32}$&  0.50&  0.25&  -7.6670& -11.4350&  0.1660  &-682.12201  
&-682.85199\\
$^{80}{\rm Ge}_{32}$&  0.25&  0.25&  -6.7390& -12.1790&  0.1530  &-690.15399  
&-690.39801\\
$^{81}{\rm Ge}_{32}$&  0.25&  0.25&  -6.3290& -12.7720&  0.1149  &-695.07599  
&-696.47198\\
\hline
\end{tabular}
\end{center}
\end{table}
\clearpage

\begin{table}[t]
\begin{center}
\caption{ 
Results of RMF calculations for various isotopes of Ni using the
parameter set NL3. The values of the fixed proton and neutron pair-gaps
($\Delta_n$ and $\Delta_p$) used in these calculations are identical
to the values used in calculations with parameter set NL-SH. The remaining
columns give the computed values of $\mu_n$, $\mu_p$, and the differences
between the calculated and experimental binding energies in table
\ref{tab4: CoNi} in MeV. Negative values mean that the calculated binding
energy is larger then the experimental binding energy.
\label{tab7: NL3}}
\begin{tabular}{ccccccc}
Isotope & $\mu_n$ & $\mu_p$ & $ BE_{\rm NL3}$ & 
$BE_{\rm NL-SH}$ \\
       &          &         & $-BE_{\rm Exp.}$ & $ -BE_{\rm Exp.}$ \\   
\hline
$^{58}{\rm Ni}_{28}$&  -11.044&  -5.692&   0.74099  &  0.56998  \\ 
$^{59}{\rm Ni}_{28}$&  -10.085&  -6.282&   0.60501  &  0.45209  \\ 
$^{60}{\rm Ni}_{28}$&  -10.041&  -6.861&   0.48603  & -1.23132  \\
$^{61}{\rm Ni}_{28}$&   -9.527&  -7.556&   0.44602  & -1.46258  \\ 
$^{62}{\rm Ni}_{28}$&   -9.375&  -8.101&   0.31500  & -2.37701  \\
$^{63}{\rm Ni}_{28}$&   -9.090&  -8.759&   0.26599  & -1.79053  \\ 
$^{64}{\rm Ni}_{28}$&   -8.873&  -9.320&  -0.04600  & -2.13013  \\ 
$^{65}{\rm Ni}_{28}$&   -8.619&  -9.954&  -0.29802  & -0.33136  \\ 
$^{66}{\rm Ni}_{28}$&   -8.479& -10.582&  -0.54602  & -2.24903  \\ 
$^{67}{\rm Ni}_{28}$&   -7.873& -11.078&  -1.09099  &  0.39599  \\
$^{68}{\rm Ni}_{28}$&   -7.133& -11.645&  -1.60199  & -0.42998  \\
\hline
\end{tabular}
\end{center}
\end{table}

\begin{table}[t]
\begin{center}
\caption{ 
Comparison of the nucleonic chemical potentials 
for RMF and El Eid \& Hillebrandt EOS in stellar collapse
using data for a star of initial mass $15 M_\odot$.
\label{tab8: RMF-EH}}
\begin{tabular}{ccccccc}
Model & $\rho_{10}$ & $Y_e$ & Mean Nucleus & $ \mu_n$ (MeV) & 
$\mu_p$ (MeV) \\
\hline
RMF & 0.37 &  0.42 &  68 & -6.93  &  -11.7  \\ 
EH  & 0.37 &  0.42 &  52 & -7.08  &  -11.5  \\ 
\hline
\end{tabular}
\end{center}
\end{table}

\begin{figure}[h]
\caption{
(a) Cumulative neutrino flux up to  $\rho_{10}=24.16$ g/cm$^3$, 
computed with Fuller/BBAL EOS  for M = 25 M$_{\odot}$, 
D = 1 kpc and $|M_{GT}|^2$ = 2.5 and 
later 0.1.
(b) The spectrum in (a) folded with the detection cross-section for
c.c. reaction $\nu_e(d,pp)e^-$ in SNO.
(c) The spectrum in (a) folded with the detection cross-section
for $\nu_e -e^-$ scattering in Super Kamioka. The projected
lower limit $E_{\nu}$ is $\geq 5$ MeV for the S-K and SNO detectors.
\label{fig1: 25msmhv2}
}
\end{figure}

\begin{figure}[h]
\caption{
The chemical potential of our RMF-calculations with the parameter
set NL-SH $\mu_n|_{RMF}$ is compared with that of the BCK-model 
$\mu_n|_{BCK}$ and with the functional forms of the liquid drop model 
(LDM) given in Eq. (\ref{eq: Fullmun}) and the finite range liquid drop 
model (FRLDM) $\mu_n|_{FRLDM}$ discussed in Sect. \ref{sec: FRLDM} for 
Mn isotopes
\label{fig2: ch5mn}
}
\end{figure}

\begin{figure}[h]
\caption{Experimental vs. calculated values of binding energies using 
the parameter set NL-SH.. The calculations were carried out with the
the pairing gap parameters listed in the Tables \ref{tab3: MnFe},
\ref{tab4: CoNi}, \ref{tab5: CuZn} and \ref{tab6: GaGe}.
\label{fig3: Ch4fig1}
}
\end{figure}

\begin{figure}[h]
\caption{Neutron chemical potentials of Ni isotopes calculated 
within RMF theory based on the parameter set NL-SH and within the 
BCK-model are compared with the functional form of $\mu_n$ in 
Eq. (\ref{eq: BBCWmun}) (solid line) and Eq. (\ref{eq: Fullmun}) 
(dashed line).
\label{fig4: BBCW}
}
\end{figure}
         
\begin{figure}[h]
\caption{The chemical potentials $\mu_n|_{RMF}$ (parameter set NL-SH) 
and $\mu_n|_{BCK}$ for Ni. The values $\mu_n|_{BCK}$ have been generated 
from the BCK-EOS using $S_v=30.34$ 
MeV but with the same incompressibility $K = 180$ and $a_v=16.0$.
\label{fig5: ch5bcknewn}
}
\end{figure}
                   
\begin{figure}[h]
\caption{
Difference of proton and neutron chemical potentials 
$\hat \mu|_{RMF}$ (parameter set NL-SH) and $\hat \mu|_{BCK}$ for Ni.
The BCK-values have been generated from the BCK-EOS using $S_v=30.34$ 
MeV, but with the same incompressibility $K = 180$ and $a_v=16.0$.
\label{fig6: ch5bcknewh}
}
\end{figure}
                   
\begin{figure}[h]
\caption{The chemical potentials $\mu_n|_{RMF}$ and $\mu_n|_{BCK}$, 
along with the functional forms of the liquid drop model (LDM) 
given in Eq. (\ref{eq: Fullmun}) and $\mu_n|_{FRLDM}$ discussed in 
Sect.\ref{sec: FRLDM} for Ni isotopes. This figure is especially 
significant because all the Ni isotopes are spherical nuclei.
\label{fig7: ch5ni}}
\end{figure}
           
\begin{figure}[h]
\caption{The chemical potential of neutrons obtained with RMF theory 
based on the parameter set NL-SH is compared with the functional 
form of the Finite Range Droplet Model (FRDM) for Ni isotopes. 
\label{fig8: frdmni}}
\end{figure}
       
\begin{figure}[h]
\caption{The chemical potentials $\mu_n|_{RMF}$ and $\mu_n|_{BCK}$, 
along with the functional forms of the liquid drop model (LDM) given in
Eq. (\ref{eq: Fullmun}) and $\mu_n|_{FRLDM}$ discussed in Sect.
\ref{sec: FRLDM} for Fe isotopes.
\label{fig9: ch5fe}}
\end{figure}
         
\begin{figure}[h]
\caption{The chemical potentials $\mu_n|_{RMF}$ and $\mu_n|_{BCK}$, 
Eq. (\ref{eq: Fullmun}) and $\mu_n|_{FRLDM}$ discussed in Sect.
\ref{sec: FRLDM} for Ga isotopes.
\label{fig10: ch5ga}}
\end{figure}
         
\begin{figure}[h]
\caption{ The nuclear binding energies using the spherical, incompressible
LDM, the FRLDM (using volume, surface, coulomb and coulomb exchange terms only)
and the experimental values for Ni isotopes.
\label{fig11: ch5nibe}}
\end{figure}
         
\begin{figure}[h]
\caption{The nuclear binding energies using the spherical, incompressible
LDM, the FRLDM (using volume, surface, coulomb and coulomb exchange terms
only) and the experimental values for Ge isotopes.
\label{fig12: ch5gebe}}
\end{figure}
      
\begin{figure}[h]
\caption{The difference of neutron and proton chemical potentials ($\hat
\mu = \mu_n -\mu_p$) for Ni isotopes, taking into account the lattice
correction, at a density of $10^{12}$ g/cm$^3$.
The figure also shows for comparison, the values of $\hat \mu|_{RMF}$
without lattice correction, and the (lattice independent) LDM value for
$\hat \mu$ from Eq. (\ref{eq: Fullmuh}).
\label{fig13: ch5nimuhlat}}
\end{figure}

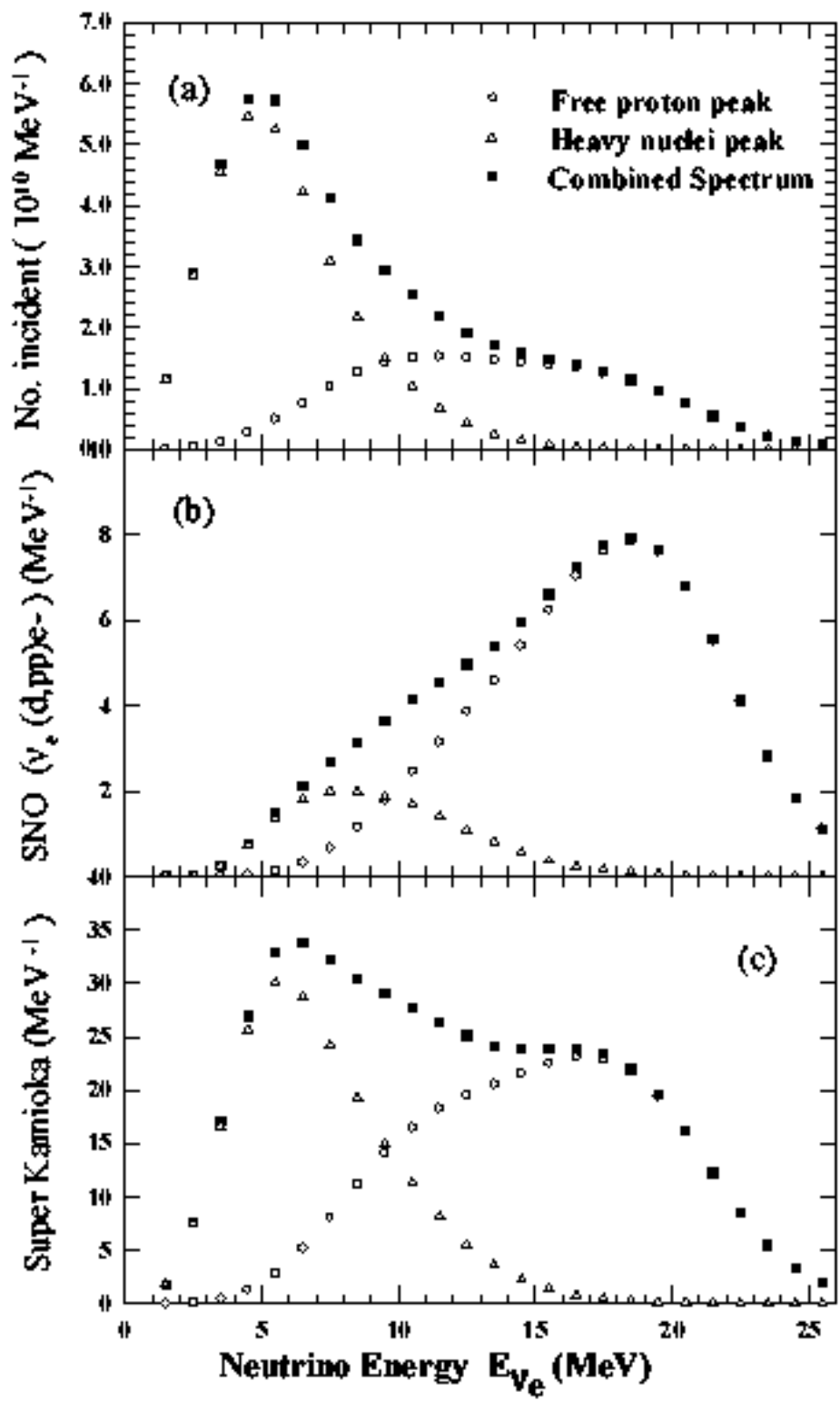
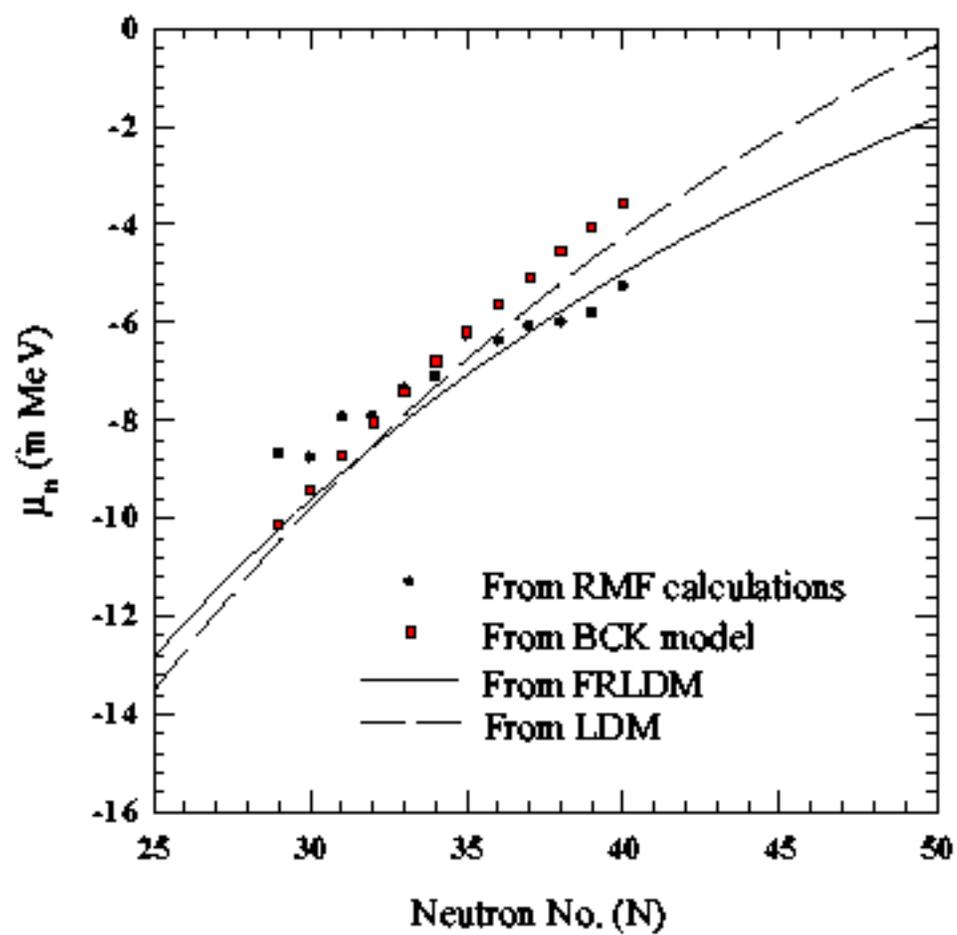
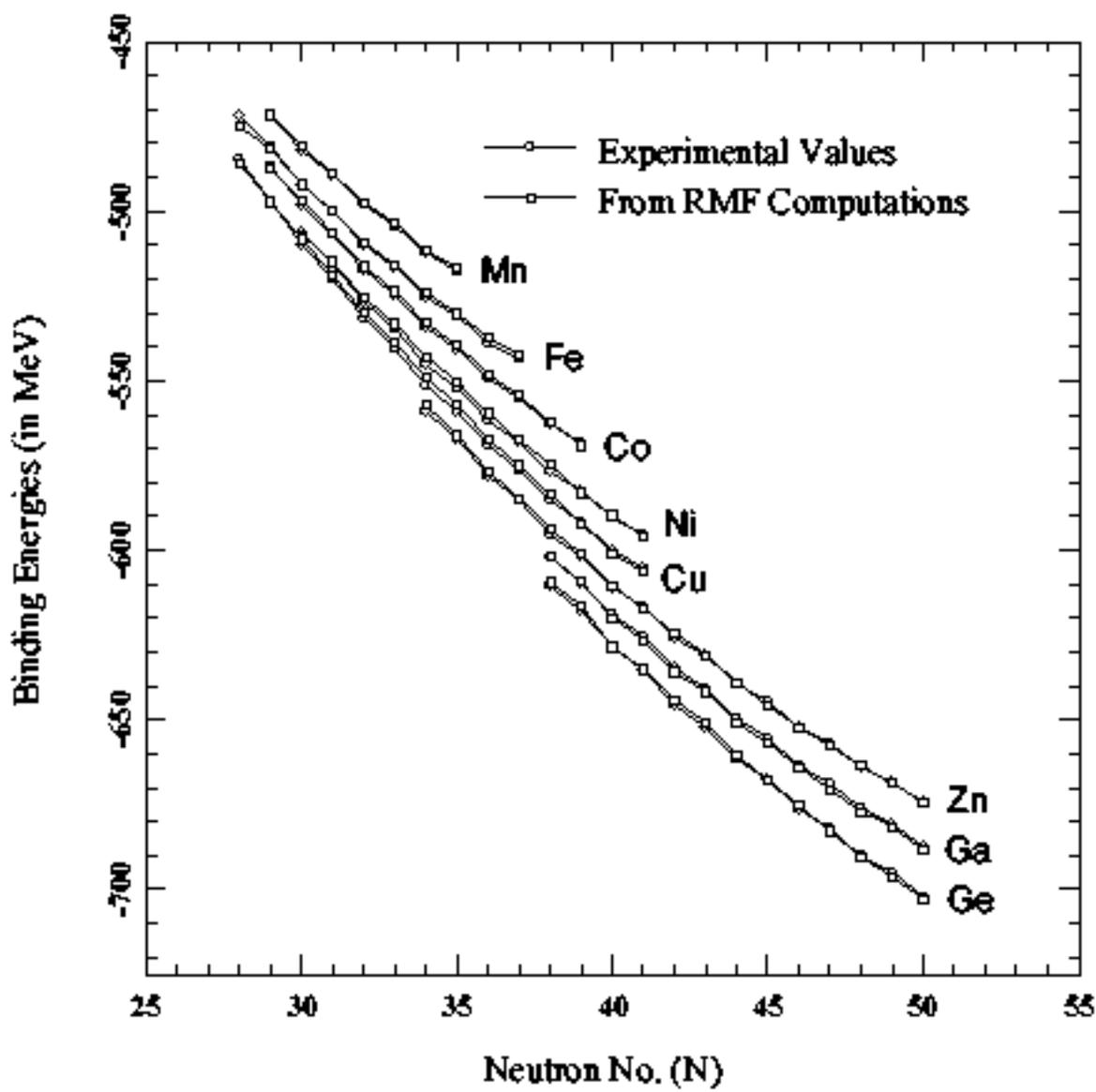
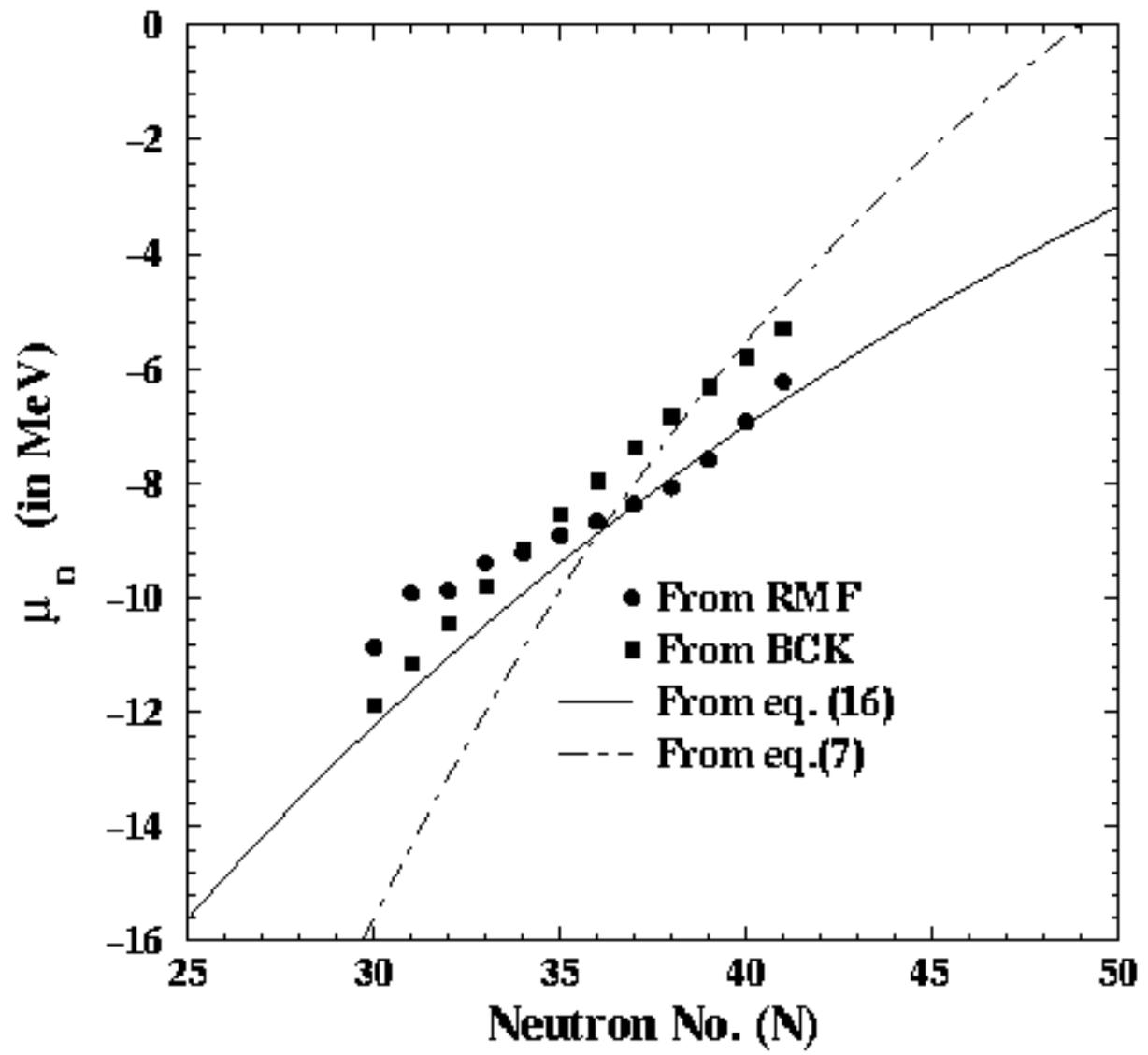
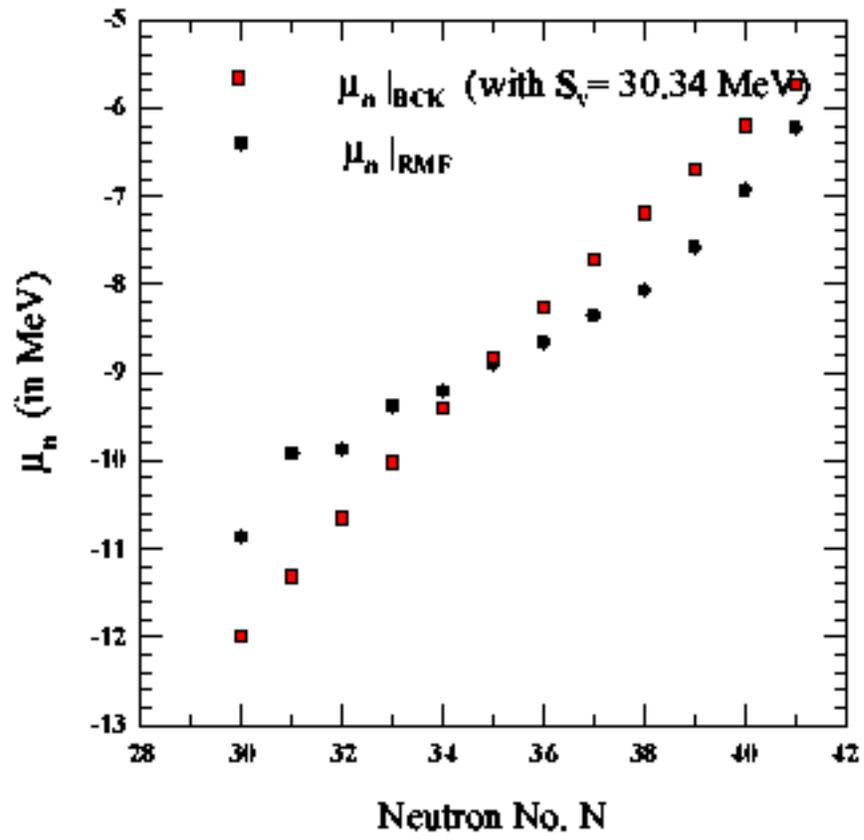
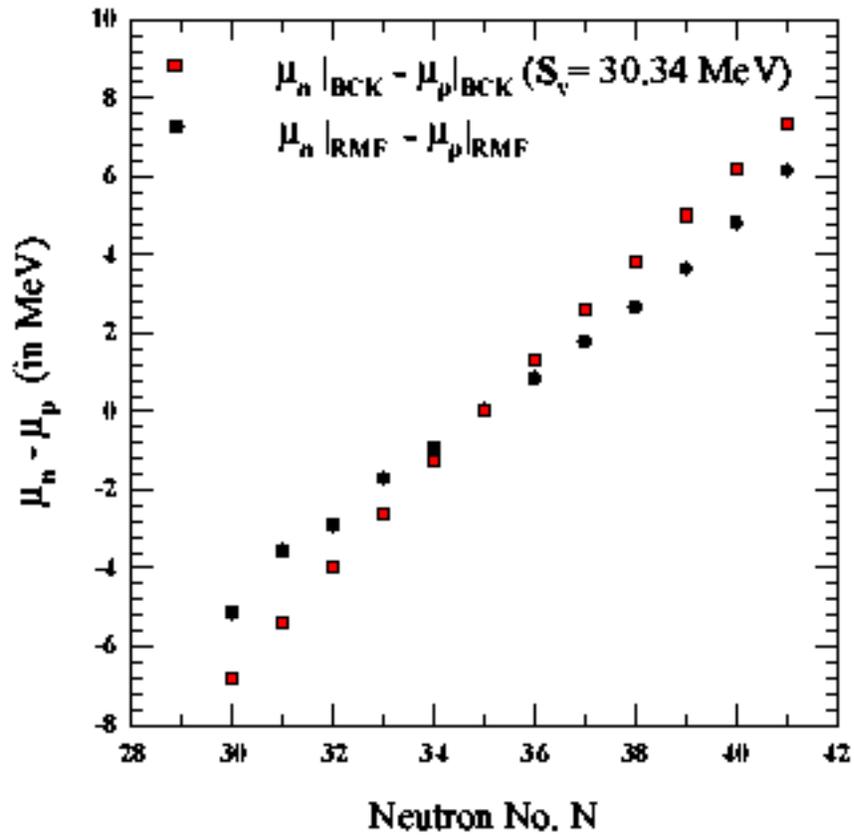
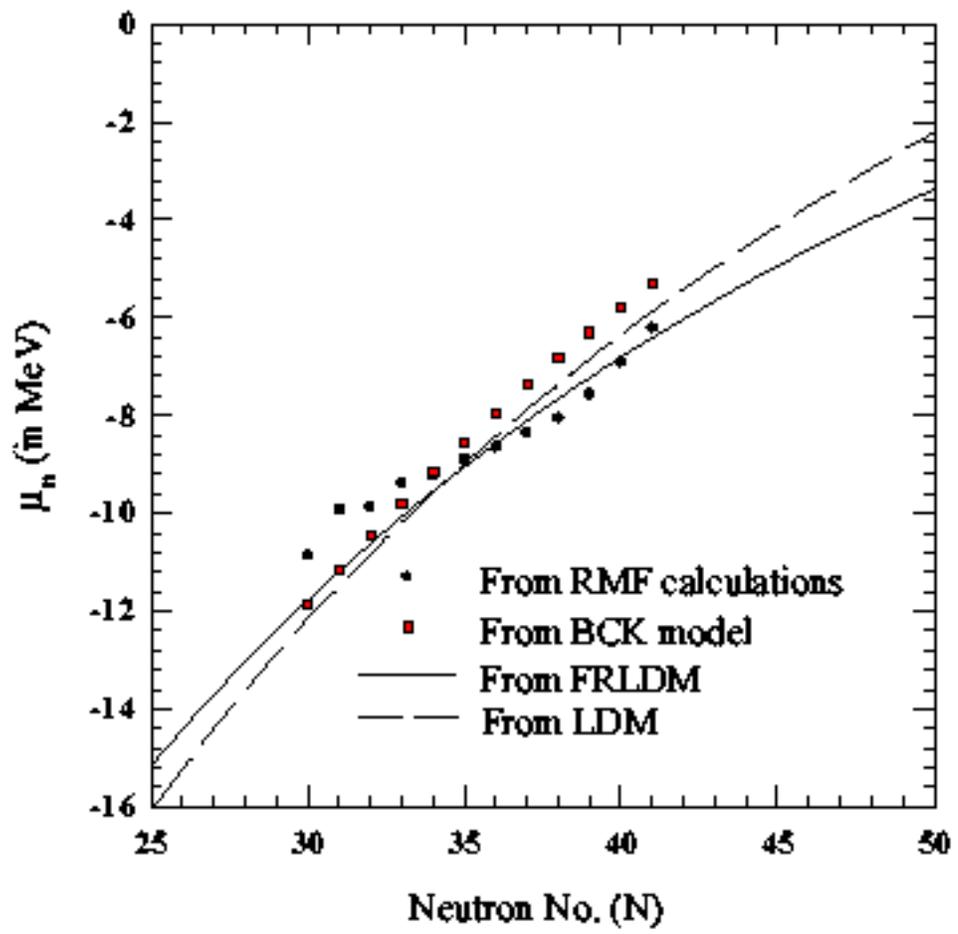
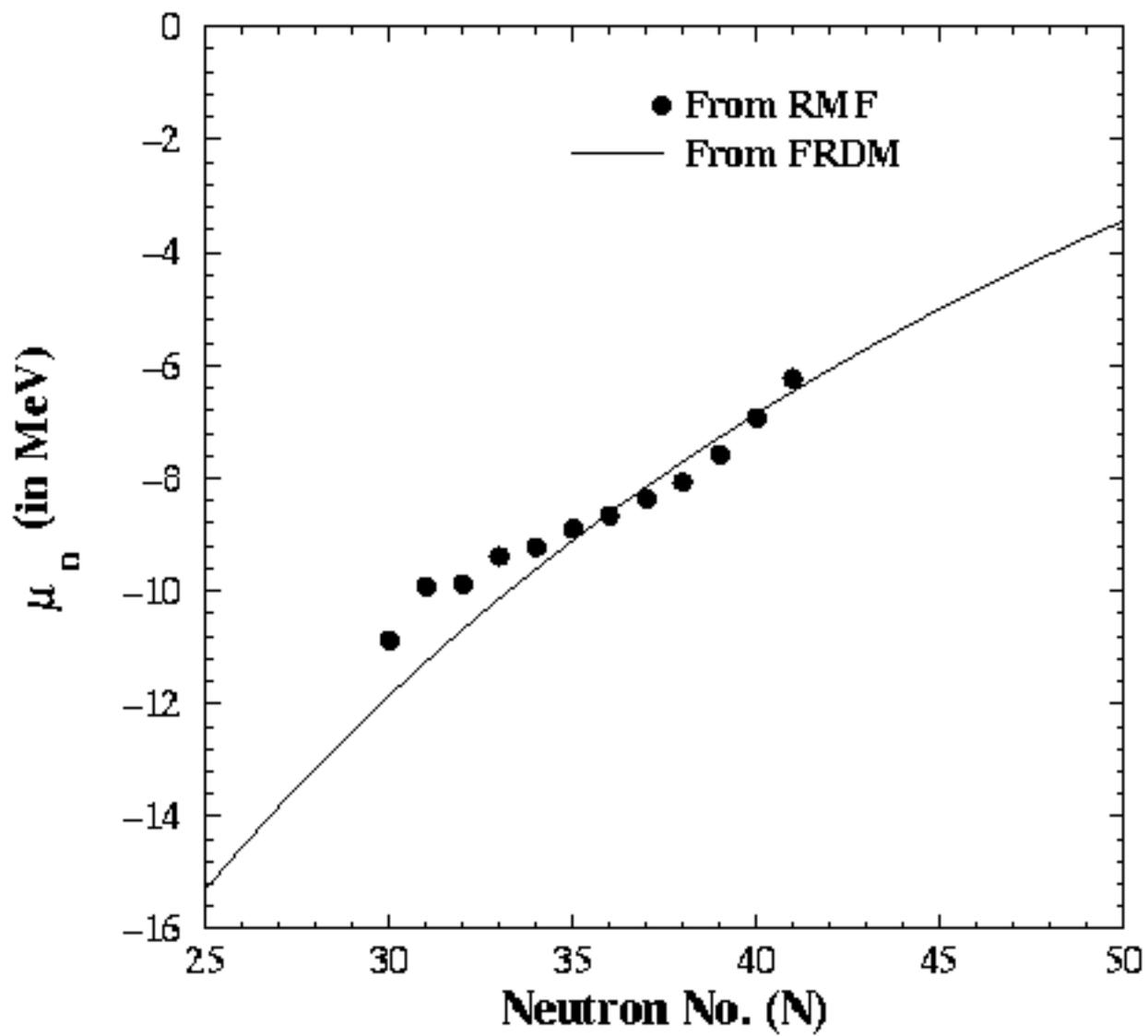
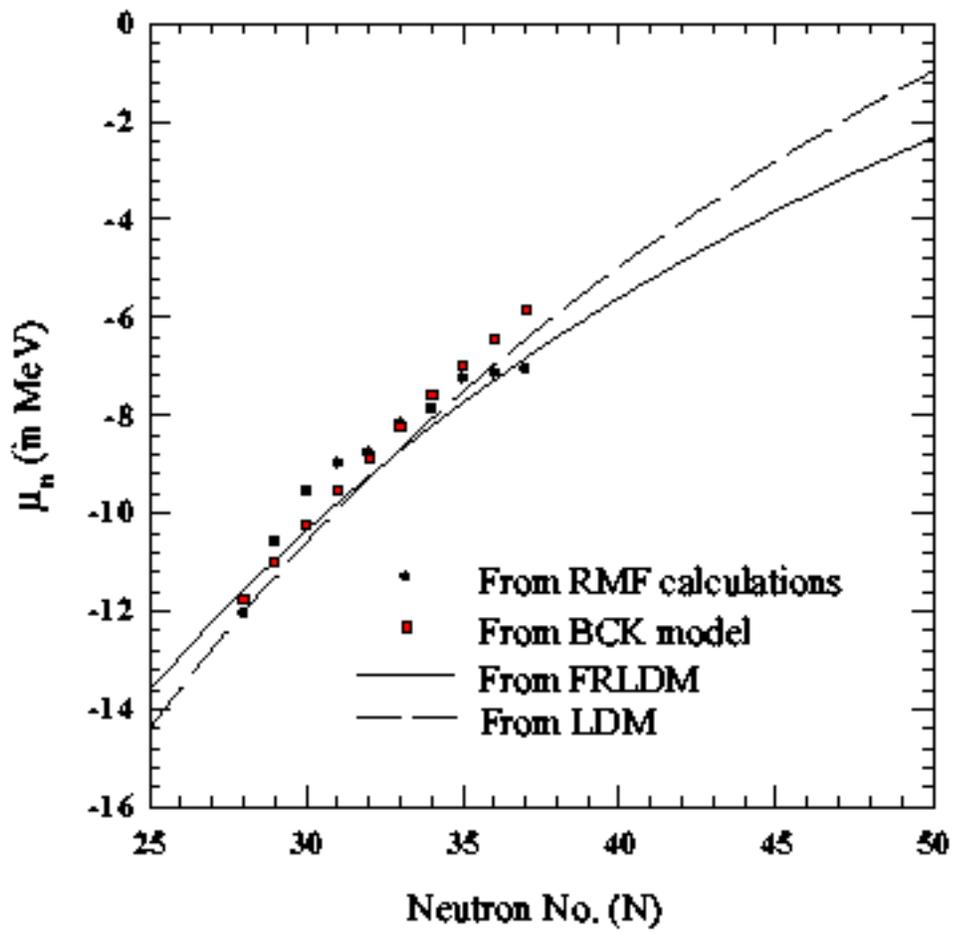
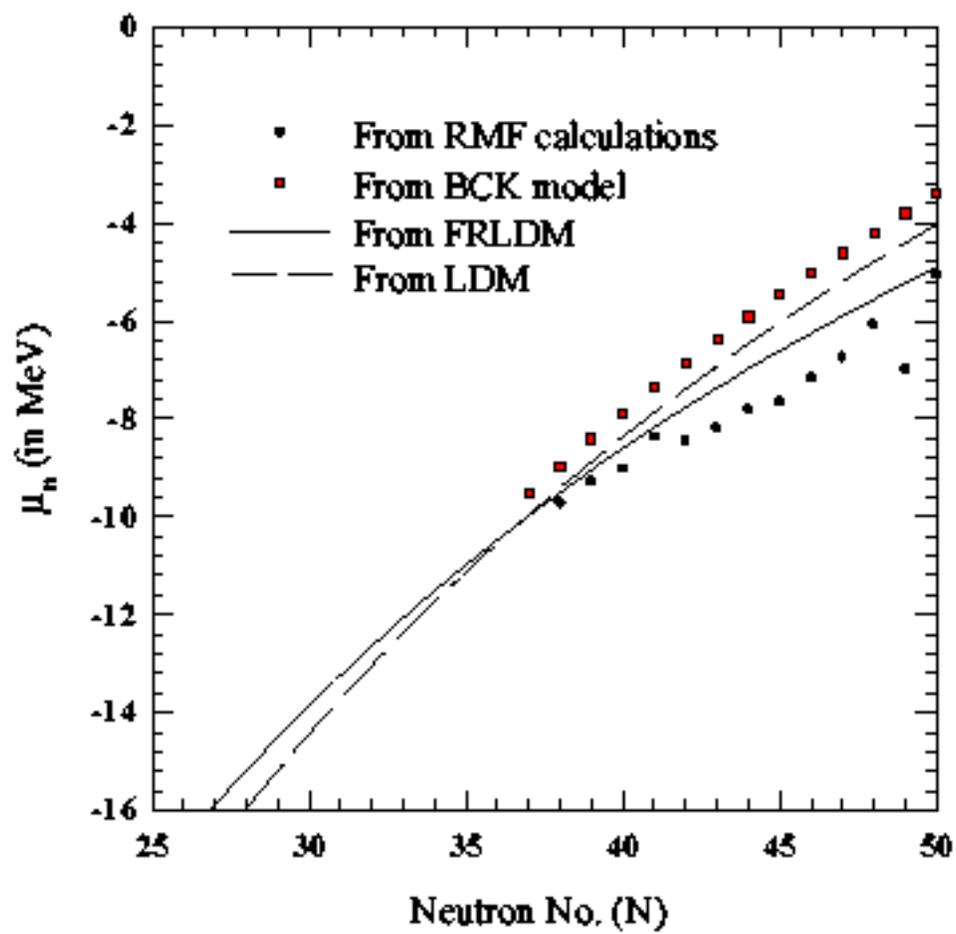
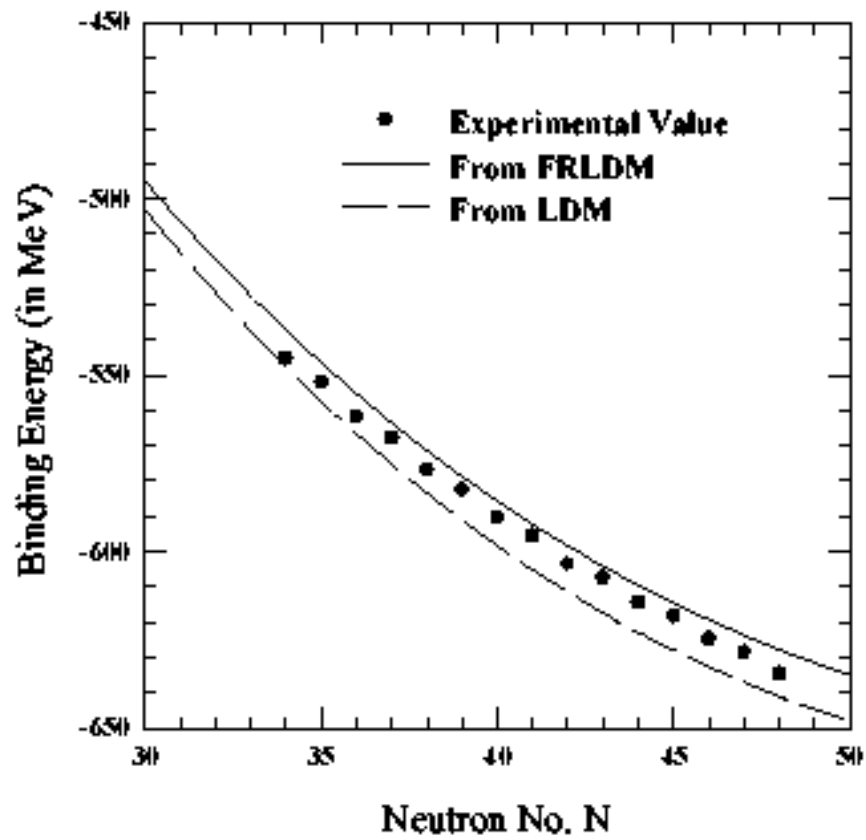
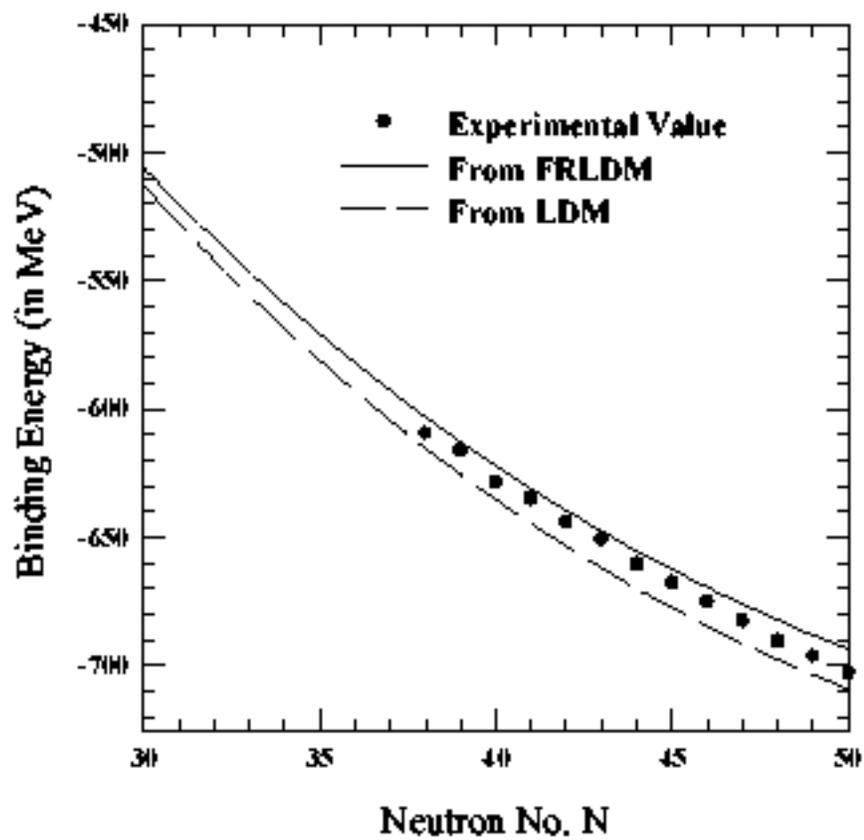
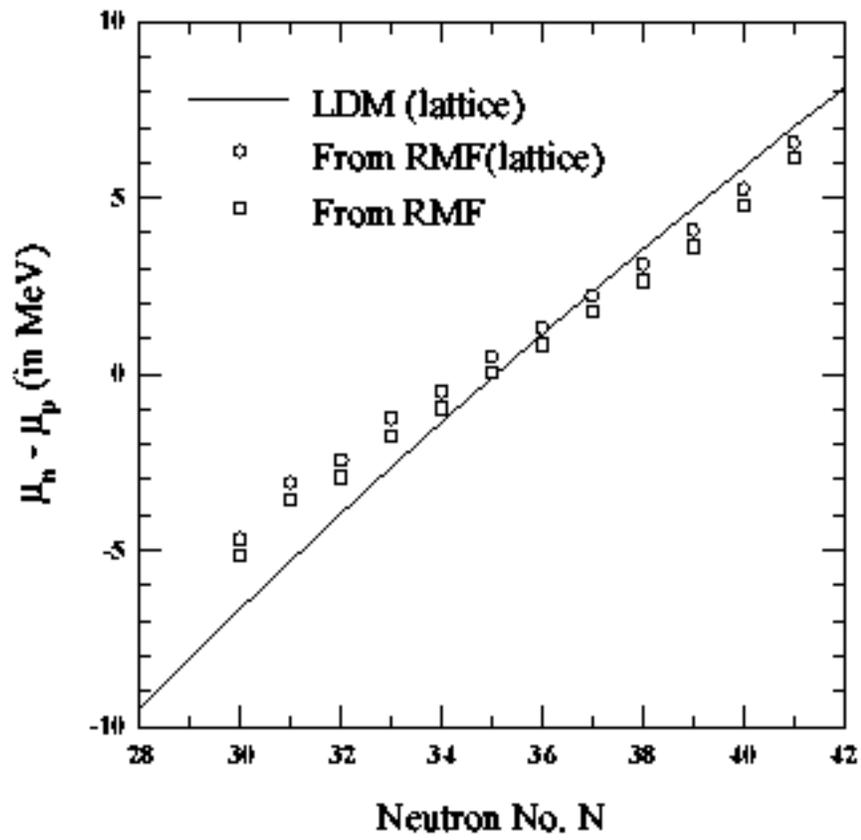
\end{document}